\documentclass[aps,prb,twocolumn,showpacs,superscriptaddress,groupedaddress]{revtex4}
\usepackage[compat=1.0.0]{tikz-feynman}
\usepackage{amssymb}
\usepackage{bm}
\usepackage{amsmath}
\usepackage{graphicx}
\usepackage{epstopdf}
\usepackage{subfigure}
\usepackage{natbib}
\usepackage{epsfig}
\usepackage{amsfonts}
\usepackage{mathrsfs}
\usepackage[toc,page,title,titletoc,header]{appendix}
\usepackage[colorlinks,linkcolor=blue,citecolor=blue,anchorcolor=blue]{hyperref}
\usepackage{dsfont,amsthm,amsbsy}

\def\be{\begin{equation}}       \def\ee{\end{equation}}
\def\bea{\begin{eqnarray}}      \def\eea{\end{eqnarray}}
\def\ba{\begin{array} }
\def\ea{\end{array} }
\def\bnum{\begin{enumerate} }
\def\enum{\end{enumerate}}

\def\=>{\Rightarrow}
\def\>{\rightarrow}

\def\eye2{Fathbb{I}}

\def\m{\textrm{matter}}

\def\eff{\mathrm{eff}}
\def\tr{\mathrm{tr}}

\begin{document}

\title{The one-dimensional Holstein model revisited}

\author{Sijia Zhao}
\affiliation{Department of Applied Physics, Stanford University, Stanford, CA 94305, USA}
\author{Zhaoyu Han}
\affiliation{Department of Physics, Stanford University, Stanford, CA 94305, USA}
\author{Ilya Esterlis}
\affiliation{Department of Physics, Harvard University, Cambridge MA 02138, USA}

\author{Steven~A.~Kivelson}
\affiliation{Department of Physics, Stanford University, Stanford, CA 94305, USA}

\date{\today}

\begin{abstract}
We analyze the global ground-state (quantum) phase diagram of the one-dimensional Holstein model at half-filling as a function of the strength of the electron-phonon coupling (represented by the strength of the phonon-induced attraction, $U$) and the phonon frequency, $\omega_0$.  In addition to reanalyzing the various asymptotic regimes, we have carried out density-matrix renormalization group  simulations to correct previous inferences concerning the anti-adiabatic (large $\omega_0$) and strong coupling (large $U$) regimes. 
There are two distinct phases - a fully gapped commensurate charge-density-wave and a spin-gapped Luther-Emery phase with a gapless charge mode - separated by a phase boundary, with a shape that reflects different microscopic physics in the weak and strong coupling limits.
\end{abstract}

\maketitle

The interaction between charge carriers and lattice vibration plays a fundamental role in strongly correlated quasi-1D materials~\cite{Peierls,Pouget,Hohenadler,Landau,Holstein,Alvermann,HKSSReview}. 
The Holstein model~\cite{Holstein}
is probably one of the simplest microscopic models of coupled electrons and phonons, which makes it an ideal platform for exact numerical methods such as the density-matrix renormalization group (DMRG)~\cite{Fehske,Ejima,Tezuka,White}, quantum Monte Carlo (QMC)~\cite{Assaad,Hohenadler2,Clay,Hardikar,HF,KM}, and other algorithms~\cite{ED1, ED2, ED3}. Surprisingly, there remain some long-standing debates, even for the one dimension (1d) Holstein model at half filling, concerning basic facts about the structure of the zero temperature ($T=0$) phase diagram, as well as discrepancies in the critical values of couplings that mark the phase boundaries  obtained with different numerical methods~\cite{Assaad}. While early studies inferred a single ordered phase for any nonzero electron-phonon coupling and finite phonon retardation~\cite{HF,Bindloss,Bakrim}, more recent
numerical results~\cite{Assaad,Clay,Hardikar,Fehske,Ejima,Bakrim2,White} have suggested the existence of a disordered phase and at least one phase
boundary.
Specifically, Hirsch and Fradkin~\cite{HF} examined the behavior of the model as a function of $\omega_0$, the bare phonon frequency, and $U$, the bipolaron binding energy which is an appropriate characterization of the electron-phonon coupling strength, both measured in units of the electron bandwidth, $4|t|$. Based on topological constraints on the nature of the phase diagram and other considerations, they speculated that the phase diagram exhibits only one phase - a fully gapped, long-range ordered charge-density-wave (CDW) phase - everywhere off these boundaries. 
They partially corroborated this conjecture with QMC studies - among the first such studies for a fermionic system.

\begin{figure}[hbt!]
    \centering
    \includegraphics[width=1.0\linewidth]{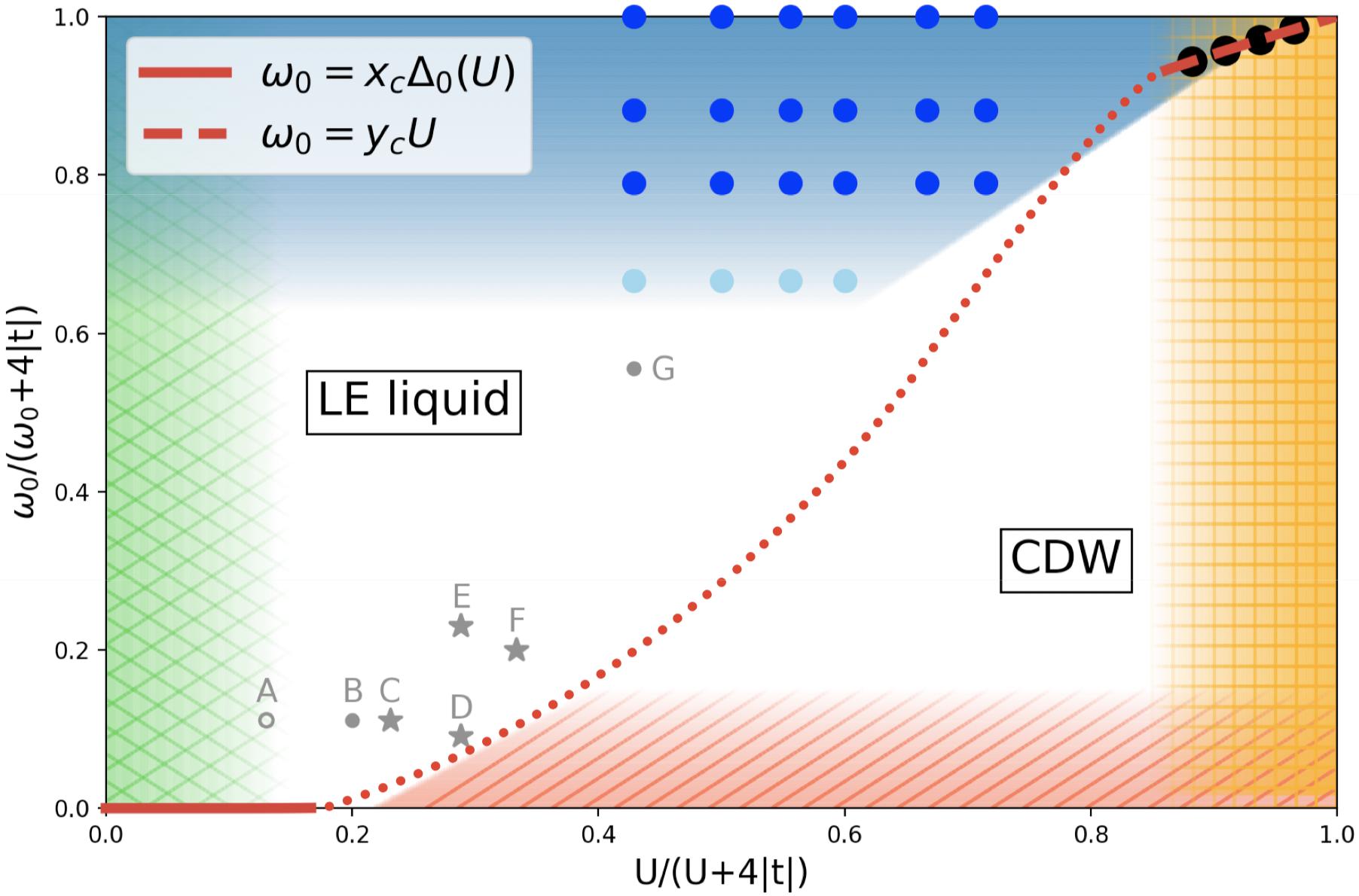}
    \caption{\label{fig:phase}
    Ground state phase diagram of the 1d Holstein model. Both phases have a spin-gap, $\Delta_s>0$.  The LE liquid phase has a gapless charge mode while the CDW is fully gapped.  
    Distinct asymptotic approaches apply at each edge of the phase diagrams: 
    (i) In the blue (anti-adiabatic) region, where $\omega_0 \gg |t|, U$, we have derived an effective Hamiltonian in powers of $t/{\omega_0}$ and $U/{\omega_0}$, and solved it using DMRG as denoted by the blue points. 
    (ii) In the orange (strong coupling) region, $U\gg |t|$, we use a combination of a strong coupling expansion~\cite{strong} and DMRG, 
    to numerically identify the position of the phase boundary, $\omega_0 \sim  y_c U$ with $y_c\approx 0.45$, 
    as indicated by the black circles.
    (iii) In the red (adiabatic) region, where $\omega_0 \ll \Delta_0$ and $|t|$ with $\Delta_0=4|t|\exp[-2\pi |t|/U]$ the mean-field  
    gap (for charge and spin), the CDW is stable against quantum fluctuations up to a critical value of $\omega_0$. For small $U\ll t$, we estimate the critical phonon frequency to be $\omega_c\sim x_c \Delta_0  $ with $x_c\approx 1$. 
    (iv) In the green (weak-coupling) region, $U\ll|t|$, a previous functional renormalization group analysis~\cite{Bakrim2} confirms an extended LE liquid phase. The solid portions of the phase boundary 
    are drawn according to the asymptotic expressions obtained in the text - the dotted portion as a conjectural smooth interpolation between these end regions.  The points labeled $\text{A}-\text{G}$ refer to 
    previous numerical studies (not our own), as discussed in the text, where those indicated by an open circle, stars, or filled circles were argued to lie in a LE liquid, a CDW, and on the phase boundary respectively.
   }
\end{figure}

In this paper, we revisit this problem and conclude that the correct quantum phase diagram of the half-filled 1D Holstein model is as shown schematically in Fig.~\ref{fig:phase}. In addition to the CDW phase, there is also a Luther-Emery 
(LE) phase, which has a
spin gap but a gapless charge mode and CDW quasi-long-range-order, i.e. it resembles an incommensurate fluctuating CDW. This structure of the phase diagram is consistent with the topological arguments of Fradkin and Hirsch in that the phase boundary does not terminate on any of the edges of the phase diagram, but rather extends from the ``corner'' at $U=\omega_0=0$ to that at $U=\omega_0 \to \infty$. 

In support of these conclusions, we have 
explored the behavior in the vicinity of the four edges of the phase diagram - each of the regions indicated by a different color of shading in Fig.~\ref{fig:phase}.  The analysis in the neighborhood of the upper ($\omega_0=\infty$) and left most ($U=0$) edges is subtle as these correspond to quantum critical lines - and of course the two corners of the phase diagram at which the phase boundary starts and ends are of particular interest:
\begin{enumerate}
    \item We have derived an effective Hamiltonian in powers of $t/\omega_0$ and $U/\omega_0$ that is valid in the vicinity of the $\omega_0 \to \infty $ (upper) edge of the phase diagram, and then solved it using high precision DMRG studies on very long (up to length $L=400$) systems. We establish that the asymptotic equivalence between the CDW and LE correlations (i.e. the emergent SU(2) symmetry as $\omega_0\to \infty$) is lifted for large but finite $\omega_0$ 
    so that there is a LE phase immediately below this edge of the phase diagram.  This is in contrast to what was conjectured by Hirsch and Fradkin, and is our most important new result.
    \item We consider a strong coupling expansion of the model - originally derived by J. K. Freericks~\cite{strong} - to fourth order in $t/U$ to explore the right edge of the phase diagram.  Again, we use high precision DMRG studies to determine the behavior of this effective model, which (as was previously known) always has an ordered CDW phase if the limit $U \to \infty$ is taken at fixed $\omega_0$.  However, we find that for  large but finite $U$, there is a phase transition from a CDW ordered state for $\omega_0 < y_c\  U$ to a LE liquid phase for $\omega_0 > y_c \ U$, where we estimate $y_c \approx 0.45$.
    \item The familiar Peierls instability ensures that for any fixed $U>0$, the ground-state is an ordered CDW in the limit $\omega_0 \to 0$, i.e. on the lower boundary of the phase diagram. 
    Specifically, for $\omega_0=0$, a mean-field analysis is exact, which predicts a finite gap $\Delta_0 = 4|t| \exp(-2\pi |t|/U)$ for both charge and spin modes.
    \item The Fermi liquid state at $U=0$ is perturbatively unstable (and in that sense is quantum critical), since weak attractive interactions inevitably lead to a state with a spin-gap. For small $U$ and $\omega_0$, we present a field-theoretic analysis that suggests that the CDW state melts with increasing $\omega_0$ at an exponentially small value, $\omega_0 = x_c\Delta_0\sim \exp(-2\pi |t|/U)$ 
    Concerning larger values of $\omega_0$, but still in this weak coupling regime, we also briefly recap a previous functional RG analysis~\cite{Bakrim2} that shows the existence of a LE liquid phase everywhere proximate to the $U\to 0$ (left) edge of the phase diagram. 
\end{enumerate}
Along the way, we comment on the relation between our results and several other numerical studies~\cite{Assaad,Clay,White} that have been carried out since the pioneering work of Fradkin and Hirsch. We also present arguments suggesting that the lightly doped system exhibits a single LE liquid phase for all $\omega_0$ and $U\neq 0$.

\section{The Model}
The Holstein model is defined as
\begin{equation}
\begin{split}
    \hat{H}=&-t\sum_{\langle ij\rangle,\sigma}\Big(\hat{c}^\dag_{i,\sigma}\hat{c}_{j,\sigma}+\text{h.c.}\Big)+\alpha\sum_{i}\hat{n}_i\hat{x}_i\\[2ex]
    &+\sum_{i}\Big[\frac{\hat{p}^2_i}{2m}+\frac{K\hat{x}^2_i}{2}\Big]
\end{split}
\end{equation}
The first term describes the hopping of electrons between nearest-neighbor sites $\langle ij\rangle$, where $\hat{c}^\dag_{i,\sigma}$ creates an electron with spin polarization $\sigma$ at site $i$. The second term describes the electron-phonon interaction, where $\hat{n}_i\equiv \sum_{\sigma}\hat{c}^\dag_{i,\sigma}\hat{c}_{i,\sigma}$ is the electron density operator and $\alpha$ is the electron-phonon coupling parameter. The last term contains the lattice degrees of freedom with $\hat{x}_i$ as an optical phonon coordinate at site $i$, and $\hat{p}_i$ as the conjugate momentum. There are three independent energy scales in this problem: electron bandwidth $4|t|$, phonon frequency $\omega_0 \equiv \sqrt{K/m}$, and an effective electron-phonon interaction strength $U\equiv \alpha^2/K$.

\section{The anti-adiabatic limit, $\omega_0 \to \infty$}
To derive an effective Hamiltonian that is valid in the $\omega_0\gg |t|, U$ limit, we perform a unitary transformation $\hat{Q}=\Pi_i \exp[i\alpha\ \hat{p}_i \hat{n}_i/K]$,
such that the transformed Hamiltonian $\hat H' = \hat{Q}^\dagger H \hat{Q}$ reads~\cite{Supplemental}

\begin{equation}\label{trans}
\begin{split}
    \hat{H'}=&-t\sum_{\langle ij \rangle,\sigma}\Big(e^{i\alpha(\hat{p}_i-\hat{p}_j)/K}\hat{c}^\dag_{i,\sigma}\hat{c}_{j,\sigma}+\text{h.c.}\Big)\\[2ex]
    &-\frac{U}{2}\sum_{i}\hat{n}^2_i+\sum_{i}\Big[\frac{\hat{p}^2_i}{2m}+\frac{K\hat{x}^2_i}{2}\Big].
\end{split}
\end{equation}
Then through direct perturbation theory up to second order, we derived the effective Hamiltonian in powers of 1/$\omega_0$ for large phonon frequency:
\begin{equation}
\begin{split}
	\hat{H}_{\eff}=&-t\sum_{\langle ij\rangle ,\sigma}\Big(\hat{c}^\dag_{i,\sigma}\hat{c}_{j,\sigma}+h.c.\Big) - \frac U2 \sum_i \hat{n}_{i}^2 \\[2ex]
    &-\frac{U}{\omega^2_0}\sum_n(\hat{j}_n-\hat{j}_{n-1})^2
	\label{eq:3}
	\end{split}
	\end{equation}
where $\hat{j_n}$ is the local current operator defined as:
	\begin{equation}
    \hat{j}_n=it\sum_{\sigma}(\hat{c}^\dag_{n,\sigma}\hat{c}_{n+1,\sigma}-\hat{c}^\dag_{n+1,\sigma}\hat{c}_{n,\sigma})
	\end{equation}
When $\omega_0=\infty$, the effective model reduces to the attractive Hubbard model, while for large but finite $\omega_0$, the leading order correction gives a finite-range effective electron-electron interaction. Higher-order corrections to $H_{\text{eff}}$ are of order $(t/\omega_0)^4$ and higher. The same effective Hamiltonian can be alternatively derived by a path integral representation. Detailed calculations are deferred to the Supplemental Material~\cite{Supplemental}.

We determined ground-state properties of this effective Hamiltonian using DMRG studies on systems up to $400$ sites long for values of $\omega_0$ between $8$ and $\infty$ and for values of $U$ between $3$ and $10$. The values explored are indicated by the blue solid circles in the phase diagram in Fig.~\ref{fig:phase}. The larger $\omega_0$ results are more reliable since this is where the effective model best approximates the original problem. All the DMRG data collected are obtained from the lowest energy state out of five trials with independently randomized initial states and all the results shown (unless otherwise stated) are extrapolated to zero truncation error, utilizing data collected with five truncation errors ranging from $1\times 10^{-7}$ to $9\times 10^{-7}$. We have checked our results do not change significantly down to truncation error $1 \times 10^{-10}$, corresponding to keeping bond dimensions up to $m=1500$. All data involving sites within $N_x/4$ to the open boundary are discarded, i.e. we only retain the data on the interval $x\in [N_x/4, 3N_x/4]$, to reduce boundary effects.

Our findings can be summarized as follows, in all cases, we conclude that the system is in a 
LE phase, characterized by a spin-gap and a single gapless charge mode. The presence of a spin-gap is inferred from the fact that the spin-spin correlation function falls exponentially with distance, as shown in Fig.~\ref{fig:S(x)}. Meanwhile, as shown in Fig.~\ref{fig:C(x)}, the existence of a gapless charge mode follows from the observation that the charge-density correlations oscillate with wave-vector $\pi$,  and have an amplitude that falls as a power of distance, i.e. as $e^{i\pi r} |r|^{-K_c}$.  The inferred values of the charge Luttinger exponent $K_c$ are shown in Fig.~\ref{fig:allKc} for all the values of $U$ and $\omega_0$ we have explored.  As expected, $K_c \to 1$ as $\omega_0 \to \infty$, independent of $U$.  Significantly, however, for $\omega_0$ large but not infinite, we find that $K_c > 1$.  This is an important consistency check, as Umklapp scattering that could stabilize a long-range ordered CDW phase is perturbatively irrelevant for $K_c >1$, but would be relevant for $K_c<1$. 

We have carried out two further consistency checks of our results.  We have computed the central charge, as shown in Fig.~\ref{fig:centralcharge}, and in all cases, we find values consistent with $c=1$ within our uncertainty.  This is the expected value for a 
LE liquid;  these results are surely inconsistent with the $c=0$ expected of a commensurate CDW with long-range order. We have also examined the nature of the state slightly away from the half-filled case.  If commensurability effects are irrelevant for $n=1$, then the system is expected to evolve continuously with doping, $\delta \equiv 1-n >0$.  Indeed, as shown in Fig.~\ref{fig:doping} and Fig.~\ref{fig:spin_doping}, we find that both the spin gap (or more precisely, the correlation length characterizing the exponential falloff of the spin correlations) and the charge Luttinger exponent evolve continuously with $\delta$. Were the system commensurate, we would expect a factor of $2$ discontinuity in the spin-gap and a jump of the Luttinger exponent to $K_c\approx 2$ for $0<\delta\ll 1$.

\subsection{{Spin-spin correlation}}\label{sec:spin}
\begin{figure}[h]
    \centering
  \includegraphics[width=1.0\linewidth]{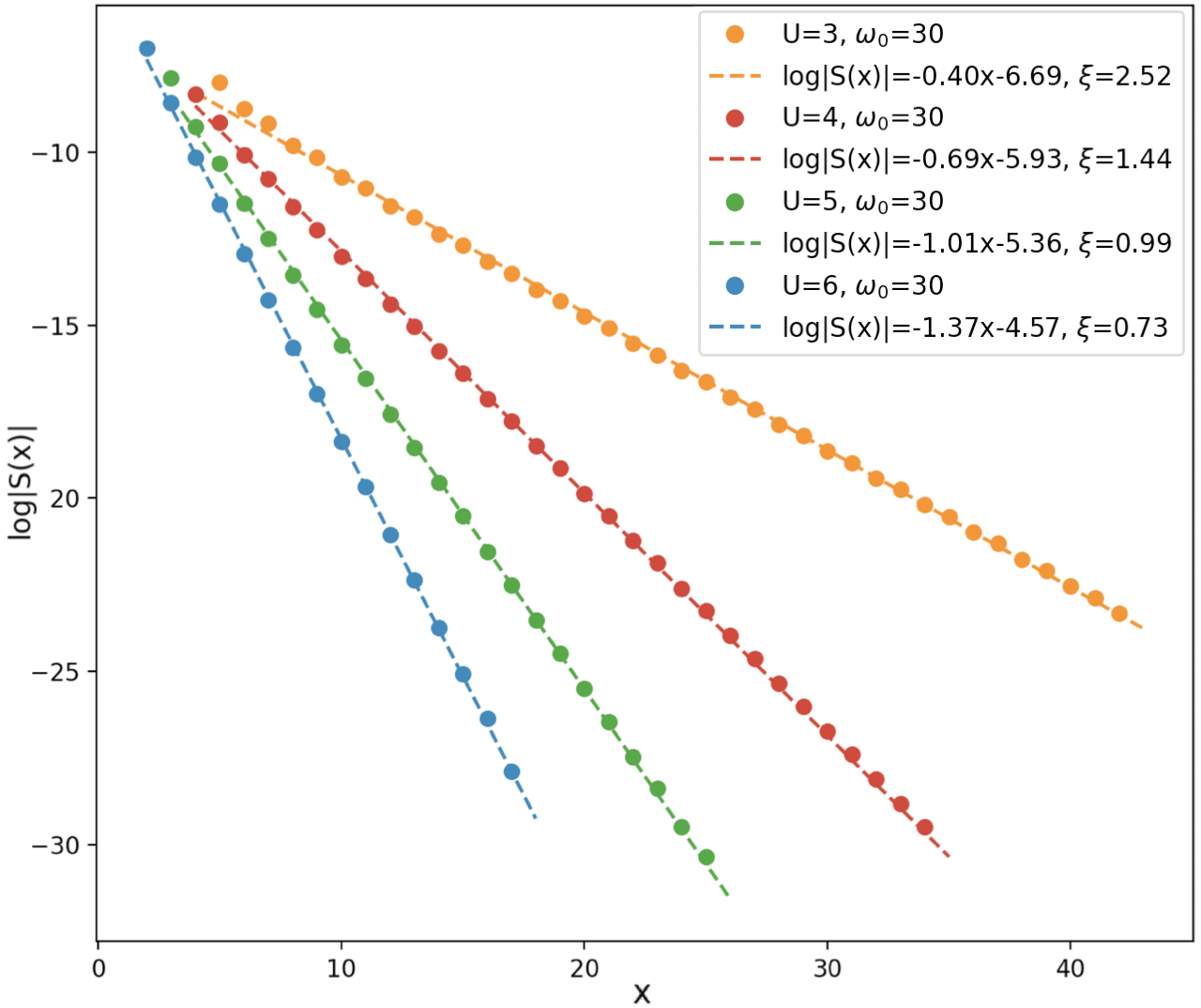}
    \caption{\label{fig:S(x)}Spin-spin correlations Eq.~\ref{spincorrelation} with exponential fit.}
\end{figure}

We have computed the spin-spin correlation function which is defined as:
\begin{equation}\label{spincorrelation}
    S(x)=\frac{1}{N_r}\sum_{r}\Big(\langle S_z(r)S_z(r+x)\rangle-\langle S_z(r)\rangle\langle S_z(r+x)\rangle\Big)
\end{equation}
where $S_z(r)$ is the $z$ component of spin operator at site $r$, and where 
we have introduced an average over $N_r=5$ ``reference sites'' near the center of the chain to reduce the finite-size effects.  As shown in Fig.~\ref{fig:S(x)}, it is clear that the spin correlators decay exponentially at large distances with a finite correlation length $\xi$ extracted by fitting the large $x$ decay of $S(x)$
 to the asymptotic  form
\be 
S(x) \sim A_S \exp[-x/\xi]\ .
\ee
The data presented in the figure are for 
$U=3,\ 4,\ 5,\ $and $6$, with $\omega_0=30$ and L=100.

\begin{figure}[h]
    \centering
  \includegraphics[width=1.0\linewidth]{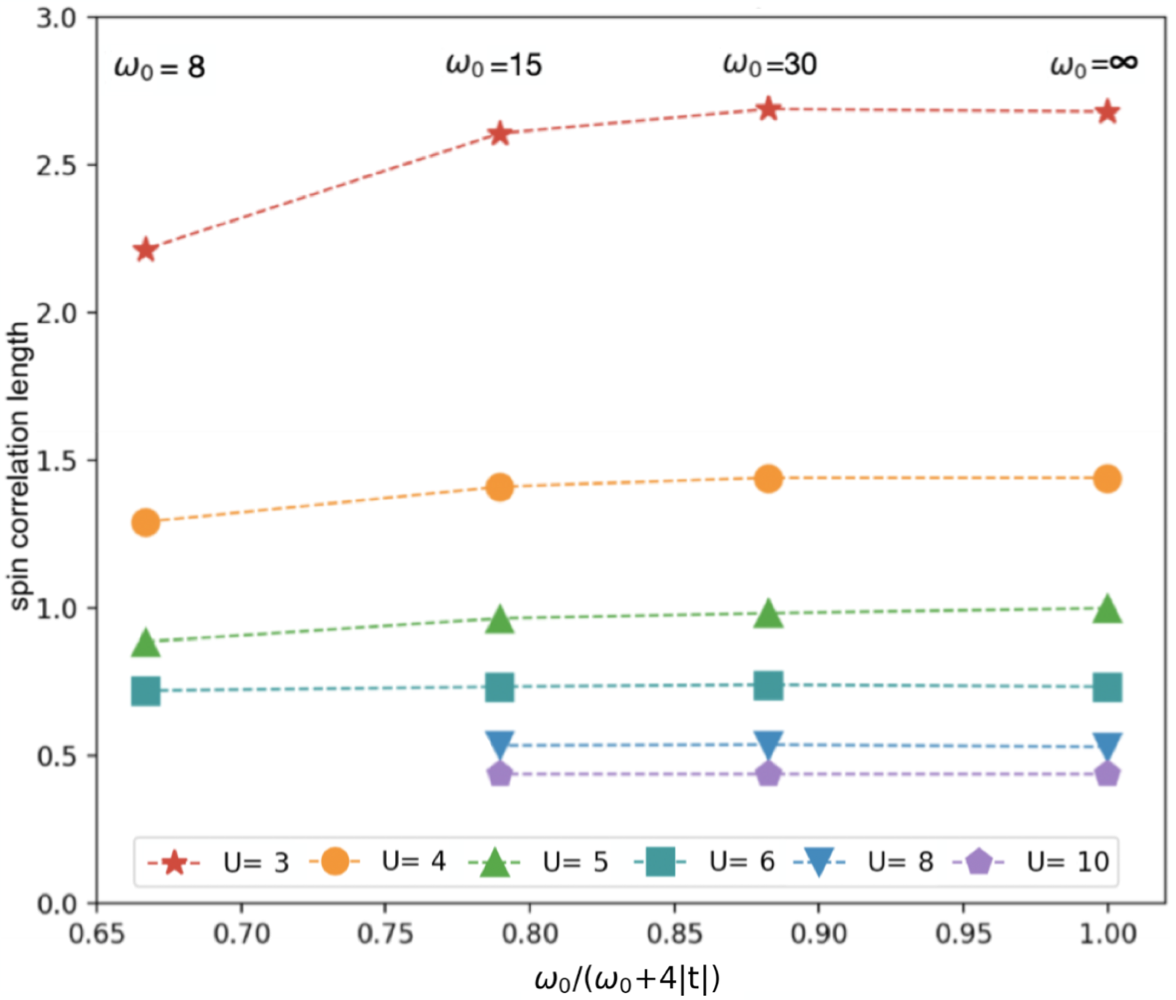}
    \caption{\label{fig:allxi}A summary of the spin correlation lengths for all values of ($U$, $\omega_0$).}
\end{figure}

That similar long-range behavior has been found for all the values of $\omega_0$ and $U$ indicated in Fig. \ref{fig:phase} confirms the non-controversial expectation that there is a spin-gap in the anti-adiabatic limit for all $U$.
A summary of $\xi$ as a function of $\omega_0$ for different values of $U$ is shown in Fig.~\ref{fig:allxi}. 

\subsection{{Density-density correlation}}
\begin{figure}[h]
    \centering
  \includegraphics[width=1.0\linewidth]{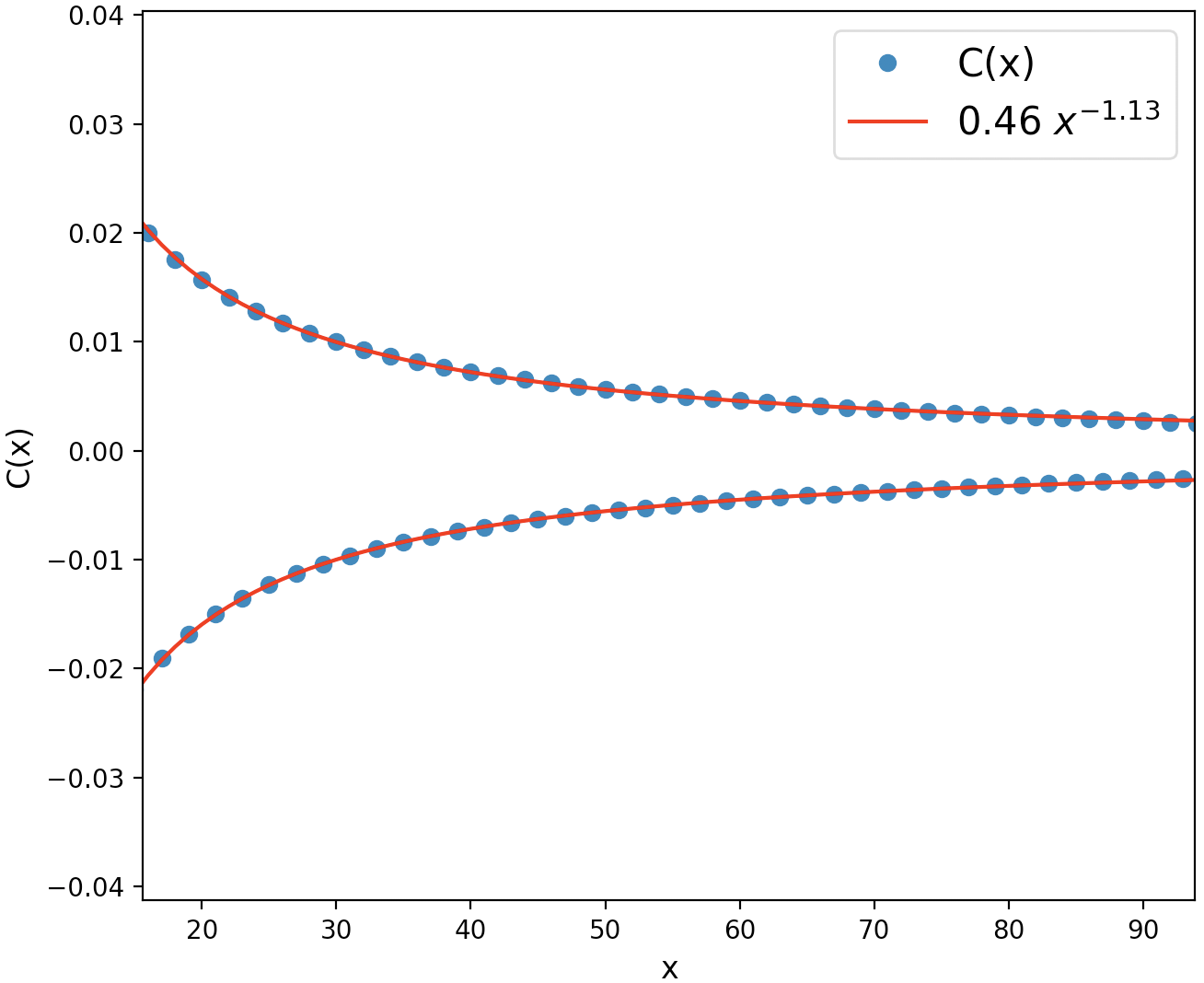}
    \caption{\label{fig:C(x)} Charge-charge correlation Eq.~\ref{chargecorrelation} for $U=6$ and $\omega_0=30$ at half filling. The Luttinger exponent is extracted using Eq.\eqref{eq:C(x)}.
    }
\end{figure}

The charge correlation function is defined as:
\begin{equation}\label{chargecorrelation}
    C(x)=\frac{1}{N_r}\sum_r\Big(\langle n(r)n(r+x)\rangle-\langle n(r)\rangle\langle n(r+x)\rangle\Big)
\end{equation}
where $n(r) \equiv \sum_{\sigma} n_\sigma(r)$ is the total density of electron on site $r$, and again we average over $N_r=5$ reference sites.
At large distance, 
we always find that $C(x)$ 
exhibits power-law behavior,
\begin{equation}
    C(x)=\frac{C_\rho}{x^2}+\frac{C'_\rho}{x^{K_C}}\cos(
    \pi x+\phi) \ ,
\label{eq:C(x)}
\end{equation}
which, from bosonization~\cite{Assaad,Giamarchi,Voit}, is the  expected behavior of a 
LE liquid with a spin gap and a charge Luttinger exponent, $K_c$.
(By contrast,  a CDW insulator with a spin gap would definitionaly exhibit long-range order at long-distances, $C(x) \sim m^2e^{i\pi x}$, where $m$ is the order parameter, and should approach this asymptotic behavior exponentially.) 
As examples of the nature of the fits to Eq.\eqref{eq:C(x)} we
have used to obtain $K_C$,
in Fig. \ref{fig:C(x)} we show the results for  ($U$, $\omega_0$)=(6, 30) 
on a chain with $L=200$. 
The dashed lines show the expected power law behavior from Eq.\eqref{eq:C(x)}
where, because we find a value of $K_c=1.13<2$, we can ignore the non-oscillatory contribution (i.e. we set $C_\rho=0$).

\begin{figure}[h]
    \centering
  \includegraphics[width=1.0\linewidth]{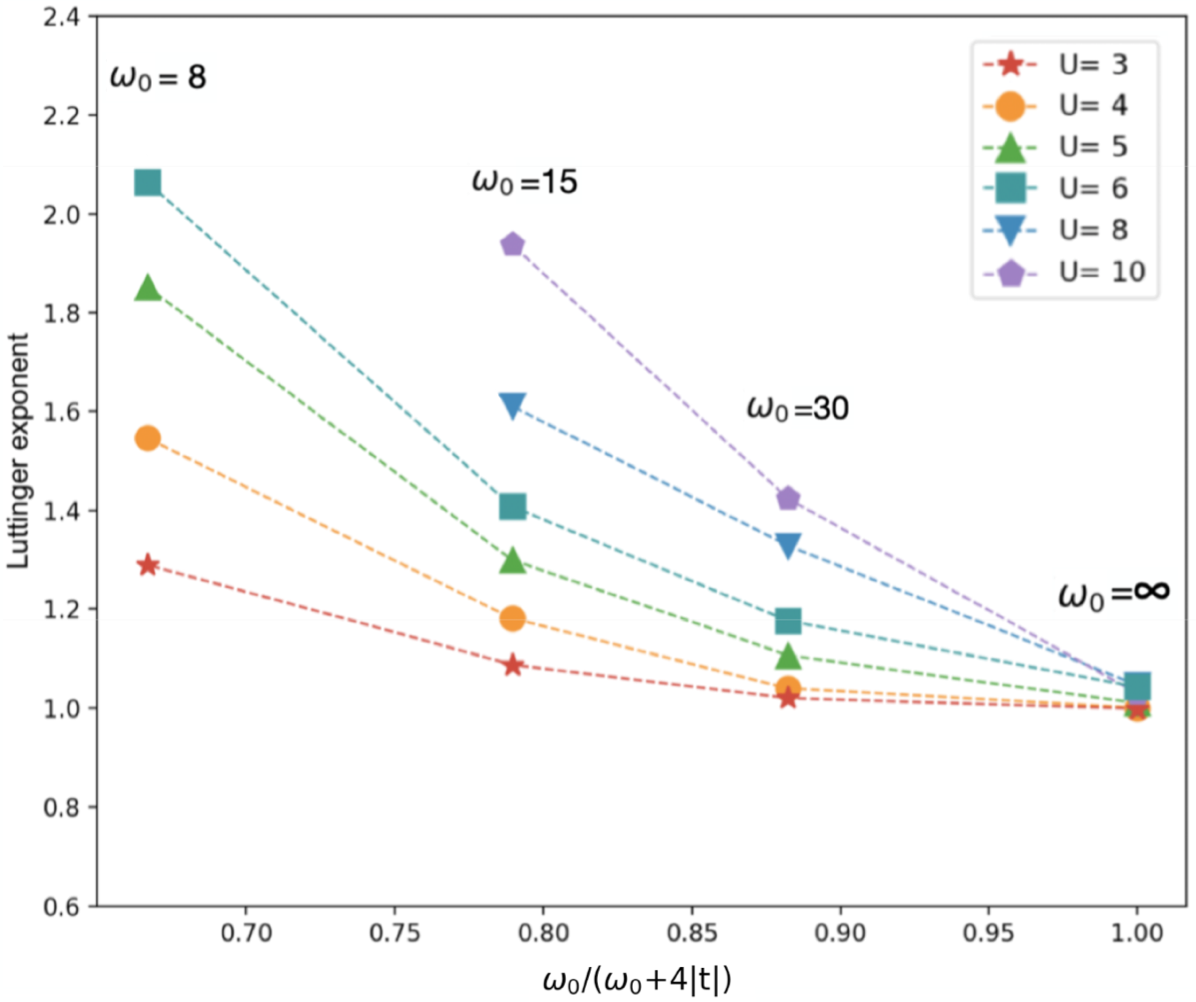}
    \caption{\label{fig:allKc}A summary of Luttinger exponents for all values of ($U$, $\omega_0$) that have been calculated by DMRG.}
\end{figure}

  The values of $K_C$ we have obtained as a function of $\omega_0$ for all 
  the values of $U$ we have considered are shown in Fig.~\ref{fig:allKc}. 
  In the limit $\omega_0\to \infty$, 
  since the Holstein model 
  maps to the attractive Hubbard model, 
  which has a charge SU(2) symmetry,
  the value of $K_C$ must approach 1, as can be seen in the figure.  
  However, for $\omega_0$ large but finite, we find $K_C>1$ for all parameters we have considered. 
  
\subsection{{Von Neumann entanglement entropy}}
\begin{figure}[h]
    \centering
  \includegraphics[width=1.0\linewidth]{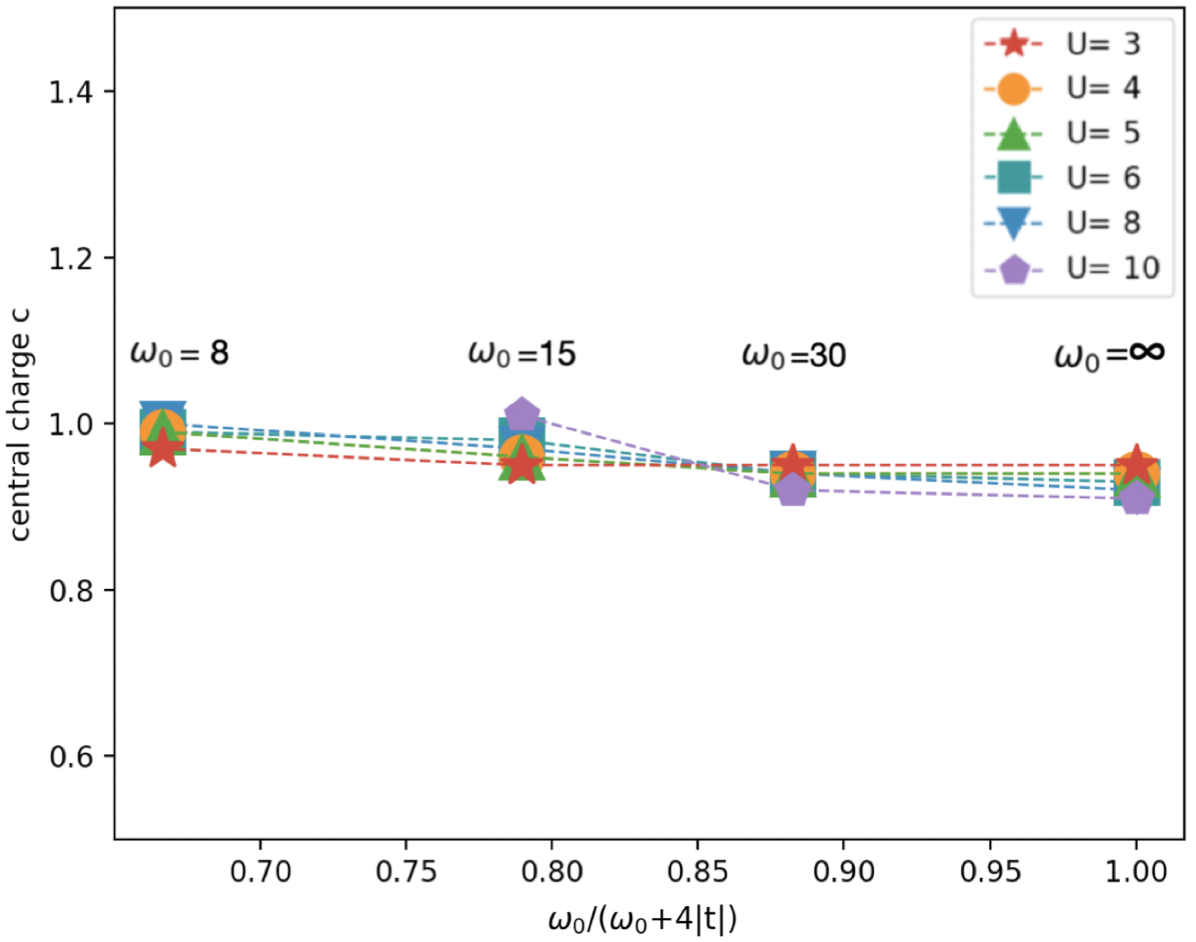}
    \caption{\label{fig:centralcharge}The extracted central charge agree well with $c=1$ for all parameter points. }
\end{figure}

\begin{figure}[h]
    \centering
  \includegraphics[width=1.0\linewidth]{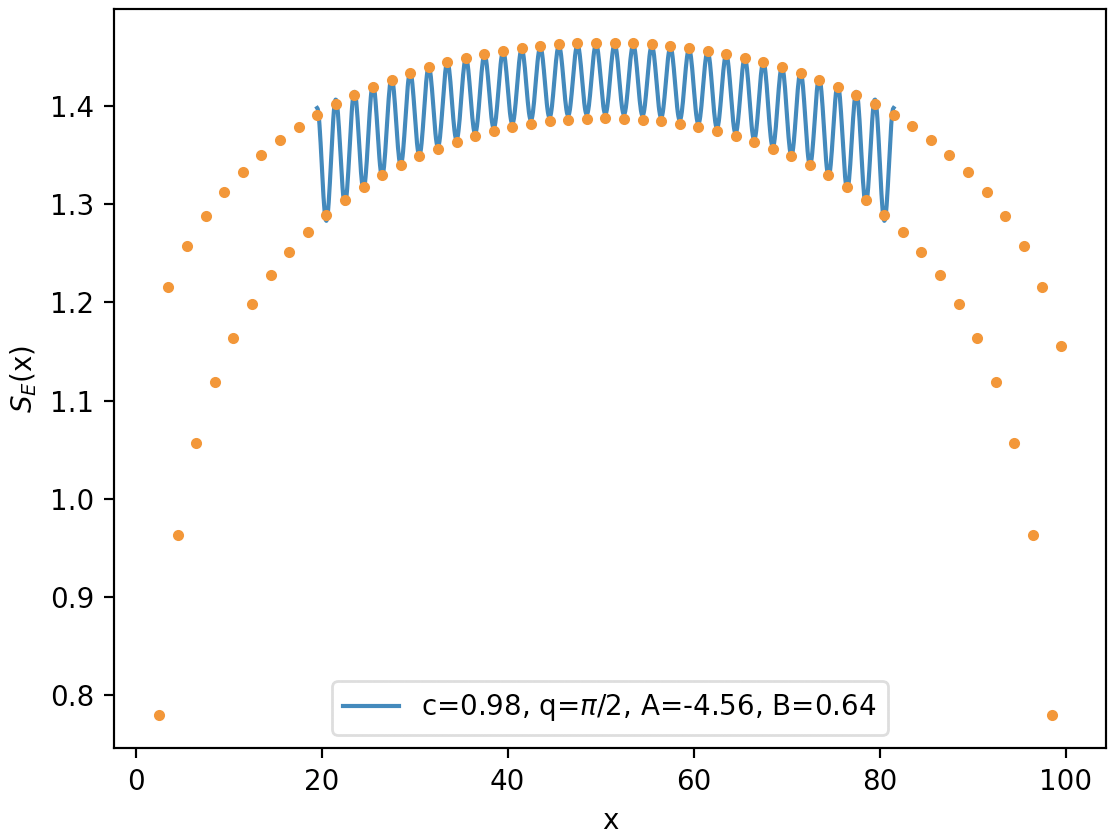}
    \caption{\label{fig:SE(x)} Here use $U=3$ and $\omega_0=8$ as an example. We fit the middle part of the system with Eq.(8) as shown by the solid blue line. The extracted values of parameters are shown in the legend. }
\end{figure}

To confirm that the system indeed 
has one gapless mode, we also calculate the von Neumann entanglement entropy $S_E(x)=-\tr(\rho_x \ln\rho_x)$, where $\rho_x$ is the reduced density matrix of a subsystem with length $x$. As has been established in \cite{entropy1,entropy2}, 
for a 1+1 dimensional system with open boundary conditions described by a conformal field theory, 
\begin{equation}
\begin{split}
    S_E(x)=&\frac{c}{6}\log\Bigg[\frac{4(N_x+1)}{\pi}\sin\Bigg(\frac{\pi(2x+1)}{2(N_x+1)}\Bigg)|\sin q|\Bigg]\\[2ex]
    &+\frac{A\sin[q(2x+1)]}{\frac{4(N_x+1)}{\pi}\sin\Big(\frac{\pi(2x+1)}{2(N_x+1)}\Big)|\sin q|} + B
\label{eq:SE(x)}
\end{split}
\end{equation}
where $N_x$ is the length of the system, and $c$, $q$, $A$, and $B$ are 
adjustable 
parameters. As expected,
we find that extrapolated to the limit $N_x \rightarrow \infty$, these fits produce a central charge, $c$, consistent with the predicted value, $c=1$, and $q=k_F$.
The quality of the fits to Eq.\eqref{eq:SE(x)} can be seen 
for representative parameters in Fig.~\ref{fig:SE(x)};  the precise values of $c$ obtained from such fits for various $\omega_0$ and $U$ are shown in Fig.~\ref{fig:centralcharge}, where we have assumed that $q=k_F$.  Within the error bars, in all cases $c=1$. 

\subsection{{Finite hole doping}}
\begin{figure}
         \includegraphics[width=1.0\linewidth]{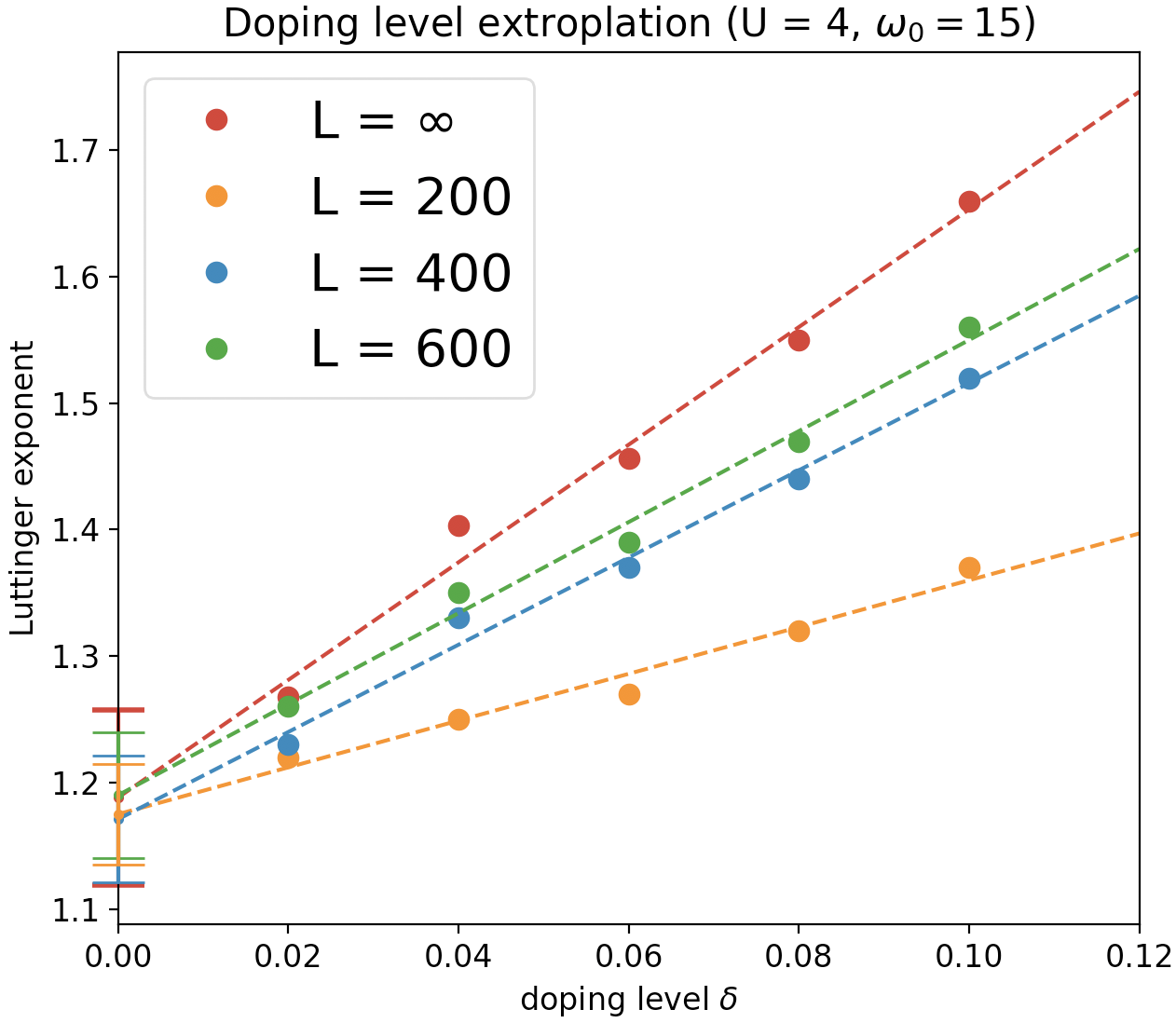}
         \includegraphics[width=1.0\linewidth]{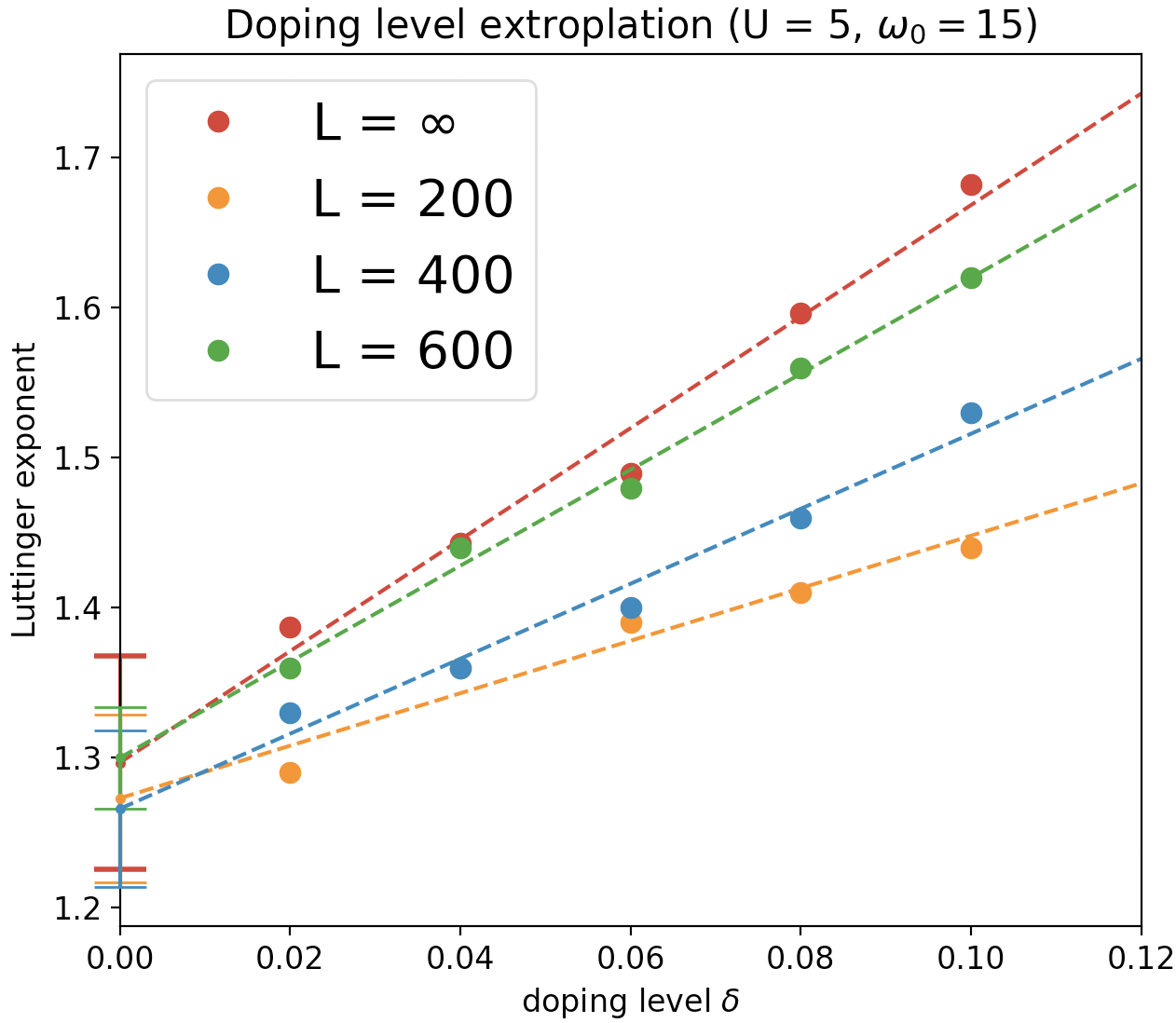}
     \caption{\label{fig:doping} $K_C$ as a function of doping level $\delta$ with error bars showing 95\% confidence bounds for the intercepts. The extrapolated $K_C$ at $\delta=0$, i.e. the intercept of the fitting function,  is 1.19 for ($U=4$, $\omega_0=15$) and 1.30 for ($U=5$, $\omega_0=15$). Both agree well with the values observed at half filling ($\delta=0$) as shown in Fig.~\ref{fig:allKc}.}
\end{figure}

We have performed one more consistency check on our numerics.  If the state at half filling is a CDW with long-range order, then upon light hole doping, $\delta\ll 1$, we generate a gas of far separated solitons.  For small $\delta$, where  these are far from each other, i.e. if $\delta \xi_{sp} \ll 1$ where $\xi_{sp}$ is the spin correlation length, the solitons should interact only through an effective hard-core interaction.  Thus they should behave like spinless fermions.  Since the system is now incommensurate, this should result in power law CDW correlations with a wave vector $Q=\pi(1+\delta)$ and with a Luttinger exponent, $K_c \to 2$ as $\delta \to 0$. The result is a discontiuity of $K_c$ at $\delta=0$, On the other hand, if the system is in a LE phase where the commensurability lock-in is irrelevant, then $K_c$ should be a continuous function of $\delta$ as $\delta \to 0$
As shown in Fig.~\ref{fig:doping}, $K_c$ 
shows no sign of a discontinuity at $\delta= 0$.

\begin{figure}[h]
    \centering
  \includegraphics[width=1.0\linewidth]{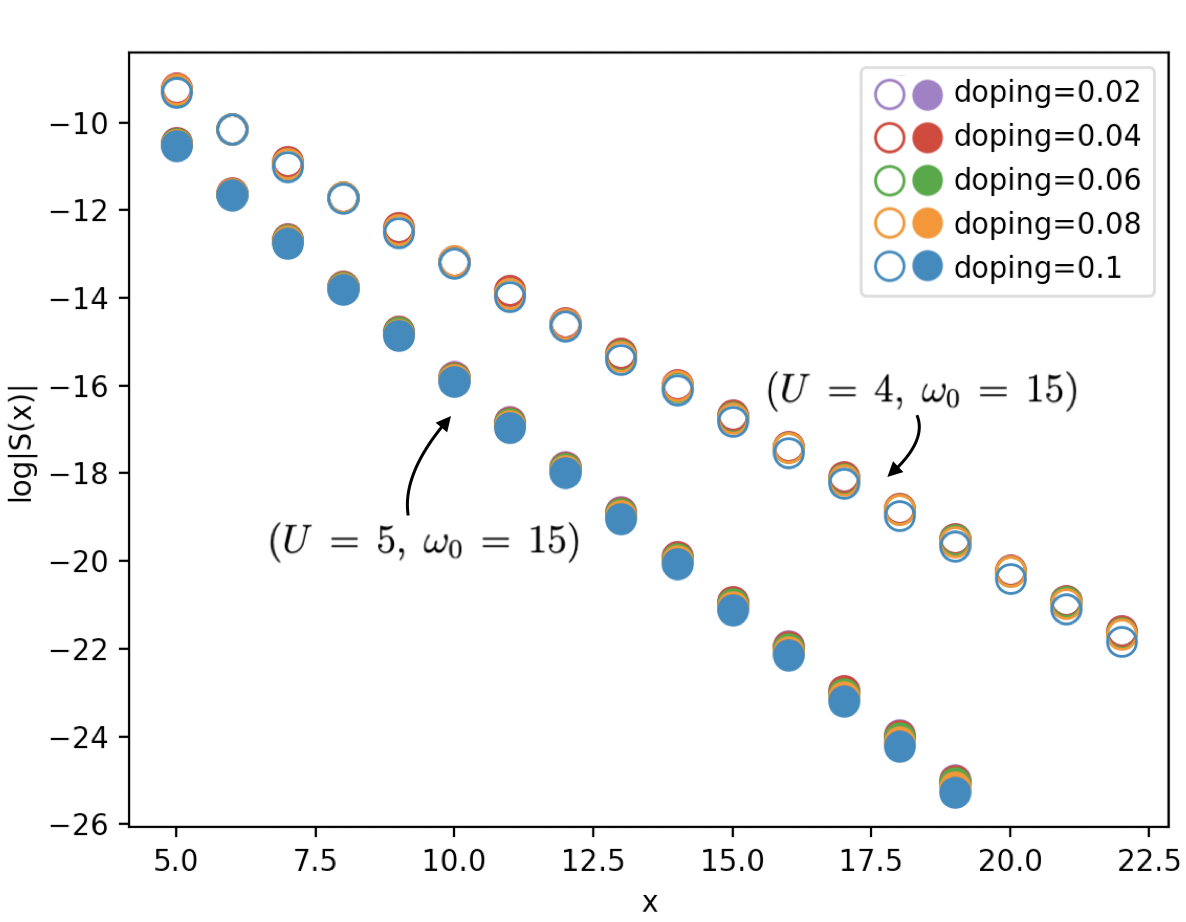}
    \caption{\label{fig:spin_doping}Spin-spin correlation for $(U, \omega_0)=(4,15)$ and $(5,15)$ at different hole doping concentration from 0.02 to 0.1. We see for both values of $(U, \omega_0)$, the spin correlations are essentially unchanged at different doping.}
\end{figure}

Moreover, the spin correlation length as shown in Fig.~\ref{fig:spin_doping} are essentially unchanged for different doping levels, which is as expected since doping makes little difference in the nature of the state in a 
LE liquid phase.

\section{The adiabatic limit, $\omega_0\to 0 $ }
For $\omega_0=0$, the phonons are static, and the problem reduces to a version of the Peierls problem, which can be exactly treated with a mean-field analysis. In other words, the ground state of the system can be obtained by optimizing the energy with varying the phonon coordinates. For all non-zero $U$, this leads to a long-range ordered, fully gapped phase with a gap $\Delta_0$ of magnitude 
$\Delta_0 \approx 4|t|\exp[-2\pi |t|/U]$ for small $U$. Moreover, it is easy to see that the CDW is stable for small non-zero $\omega_0$ so long as $\omega_0 \ll \Delta_0$.

\section{The weak coupling limit, $U\to 0$}
\subsection{The TLM model}
For small $U$, the low energy properties of the Holstein model can be characterized by an effective field theory (the TLM model)~\cite{Supplemental}.  Importantly, this effective field theory can be extended to the case of small but finite $\omega_0$, where it is identical to that which arises from the Su-Schrieffer-Heeger model.  Thus, the phase diagram must be the same in this range of parameters for the two models.  An estimate of the phase boundary in this region can be made as follows:
i) Because the model is asymptotically free, the UV cutoff can be taken to infinity in such a way that the low-energy properties are independent of it. Therefore, the soliton creation energy, which is the energy to produce an incommensuration in the CDW order, can be expressed as 
\be
E_S = \Delta_0\ F(\omega_0/\Delta_0),
\ee
independent of the cutoff energy (bandwidth).  ii) While the full form of the scaling function, $F$, is not known, the first two terms for small argument have been computed \cite{HKSSReview,NM1,NM2}, 
\be
F(x) = \frac {2}{\pi} - A x
+ {\cal O}\left( 
x^2 \right) 
\ee
where $A \approx 0.6$.
iii)  Quantum melting of the CDW order is expected to occur with increasing $\omega_0$ at the critical point, 
\be
\omega_0 = x_c \Delta_0,
\label{phaseboundary}
\ee where $F(x_c)=0$.  In other words, this is the point at which a quantum-fluctuation-driven commensurate-to-incommensurate transition occurs.  

Thus, Eq. \ref{phaseboundary} defines the phase boundary
between the LE and the CDW phases in the lower left corner of the phase diagram, where  $U\to 0$ and $\omega_0 \to 0$.  In other words, the phase boundary approaches this corner as
  \be
  \omega_0=4 x_c |t|\exp[ - 2\pi |t|/U] \ .
  \ee
Moreover, we can estimate $x_c$ 
from the first two terms in the small $x$ expansion of $F_s$ which gives $x_c \approx 2/(0.6\pi)\approx 1$.

\subsection{The functional RG method}
In Refs.~\cite{Bakrim2}, the weak coupling limit of this problem was analyzed using a perturbative RG method~\cite{twostep,RG}, which consists of successive integration of electron momentum degrees of freedom for all Matsubara frequencies divided into multiple patches. Consistent with our proposed phase diagram, it is found that as for weak enough $U$, the system flows toward a LE fixed point~\cite{Bakrim2} characterized by
a gap in the spin sector but not in the charge sector. 
To the best of our understanding, the perturbative RG is only controlled for asymptotically weak $U$.  We thus mention, but do not further analyze the fact that when the same analysis is carried out for a range of $U$, it is found that 
for fixed $\omega_0$, when $U$ exceeds a non-vanishing critical value, the Umklapp scattering becomes relevant, suggesting a transition to a phase with CDW long-range order.

\section{The strong-coupling $U\to\infty$ limit}
When the bipolaron binding energy is much larger than the electron energy scale $(|U|\gg |t|)$, performing a strong-coupling expansion for the Holstein model with the transformed Hamiltonian Eq.~\ref{trans} up to fourth order yields an effective (pseudospin) Hamiltonian~\cite{strong}:
\begin{equation}
\begin{split}
    H_{\eff}=&\sum_{i}\Bigg[t_1\Big(J^+_iJ^-_{i+1}+J^-_iJ^+_{i+1}\Big)+t_2\Big(J^+_iJ^-_{i+2}+J^-_iJ^+_{i+2}\Big) \\[1ex]
    &+2V_1\Big(J^z_iJ^z_{i+1}-\frac{1}{4}\Big)+2V_2\Big(J^z_iJ^z_{i+2}-\frac{1}{4}\Big)\Bigg]
    \label{eq:tv}
\end{split}
\end{equation}
where 
\begin{equation}
\begin{split}
    J^+_j=(-1)^jc^+_{j\uparrow}c^+_{j\downarrow},~J^-_j=(J^+_j)^\dagger,~J^z_j=\frac{1}{2}(n_{j\uparrow}+n_{j\downarrow}-1)
    \label{eq:pseudospin}
\end{split}
\end{equation}
These pseudospin operators satisfy an SU(2) algebra and form a spin-$\frac{1}{2}$ representation, where a doubly occupied site corresponds to an up pseudospin, and an empty site corresponds to a down pseudospin \cite{strong}.

\begin{figure}[h]
    \centering
  \includegraphics[width=1.0\linewidth]{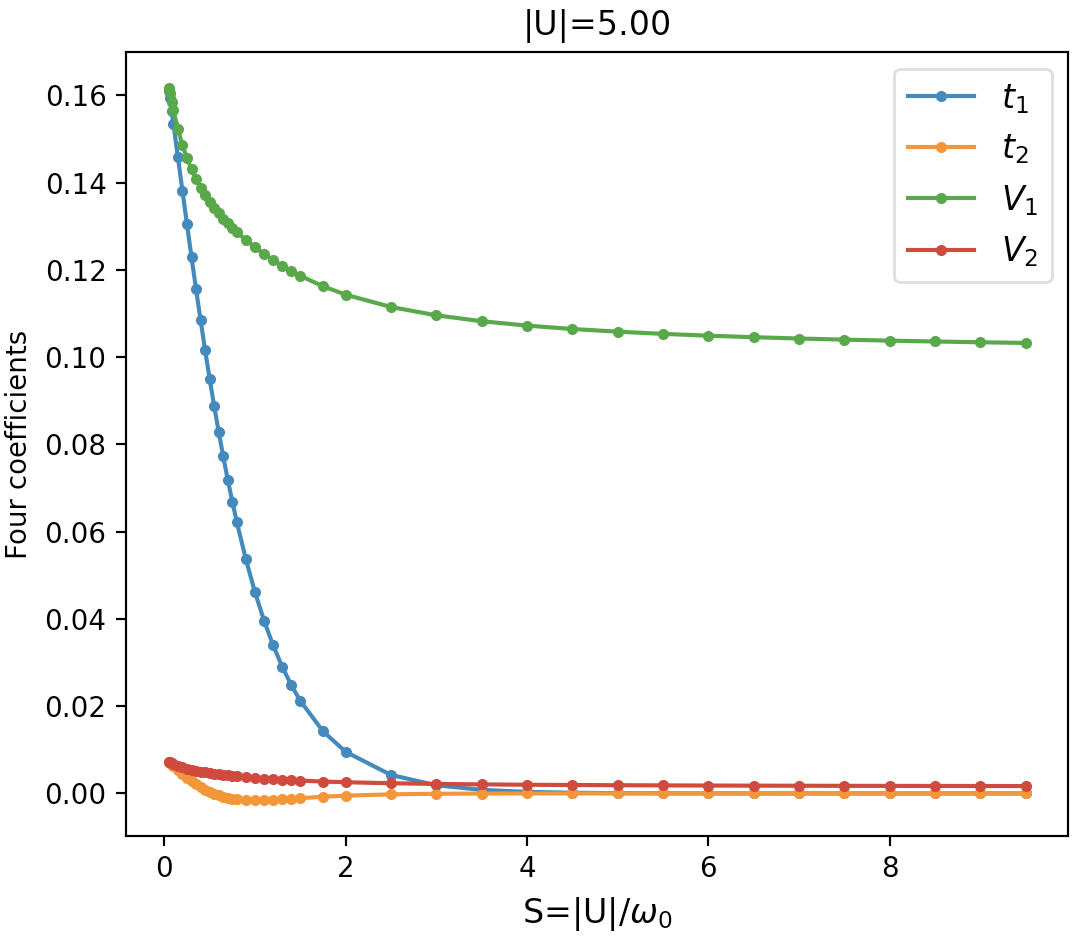}
    \caption{\label{fig:tV}An illustration of $t_1,~t_2,~V_1,~V_2$ as functions of the polaron band narrowing parameter $S=|U|/\omega_0$, with $t=1$. Here we use $|U|=5$ as an example.}
\end{figure}

In this expansion, the combination $t$ and $U$ comes out as the overall energy scale and  the only tuning parameter is the 
dimensionless retardation factor $S\equiv U/\omega_0$. In Fig.~\ref{fig:tV} we show the coefficients $t_1,~t_2,~V_1,~V_2$ as functions of $S$ for a given value of $U$ and $t$. Explicit expressions and detailed evaluation of all coefficients are given in the Supplemental Material~\cite{Supplemental}. In the anti-adiabatic limit $(S \to 0)$, these values agree with those in the strong coupling expansion of the attractive Hubbard model:
\begin{equation}
\begin{split}
    &t_1\xrightarrow{S \to 0}\ \frac{1}{4}\Big(\ \frac{4t^2}{|U|}-\frac{16t^4}{|U|^3}\ \Big)\\[1ex]
    &t_2\xrightarrow{S \to 0}\ \frac{1}{4}\frac{4t^4}{|U|^3} \\[2ex]
    & V_1\xrightarrow{S \to 0}\ t_1 \\[3ex]
    &V_2\xrightarrow{S \to 0}\ t_2.
    \label{eq:tVj}
\end{split}
\end{equation}

\begin{figure}[hbt!]
    \centering
  \includegraphics[width=1.0\linewidth]{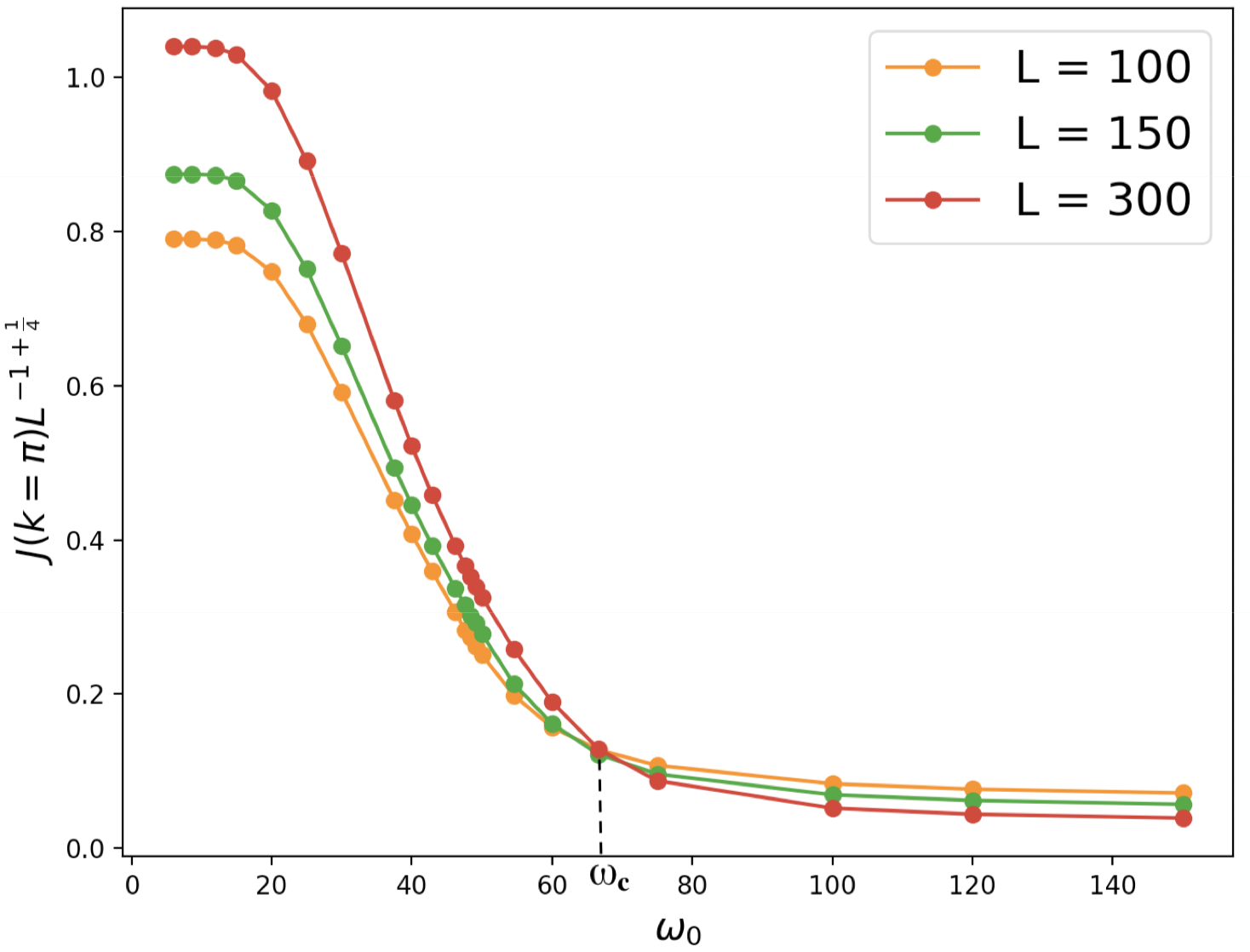}
    \caption{\label{fig:Sq_PT}  Determining the position of the phase boundary at strong coupling:  (We use data for $|U|=30$ for
    illustrative purposes.)  
    The finite size scaling properties of the structure factor $J(k)=\sum_x e^{ikx}J(x)$ evaluated at $k=\pi$ are used to  identify the critical value of $\omega_0$, where $J(x)$ is the pseudospin correlation defined as Eq.\eqref{eq:Sq}.  In the ordered phase, $J(k) \sim M^2 L^2$, while in the disordered phase $J(k) \sim L^{1-\eta}$ where $\eta > 1/4$ such that $\eta \to 1/4$ upon approach to the KT transition.  The clear crossing point in this plot establishes the existence of a KT transition between the CDW and LE phases 
    with an estimated value of the critical $\omega_0\approx 67$.
    }
\end{figure}

In the opposite limit $S \to \infty $, only $V_1$ remains non-zero and we obtain classical lattice gas, which has a CDW ground state as expected. With the coefficients determined, we then solve the effective pseudospin Hamiltonian Eq.\eqref{eq:tv} with DMRG and measure the spin-spin correlation function and the structure factor at $k=\pi$:
\begin{equation}
\begin{split}
    &J(x)=\frac{1}{N_r}\sum_{r}\ \langle J^z(r)J^z(r+x)\rangle\\[2ex]
    &J(k=\pi)=\sum_x e^{i\pi x}J(x)=\sum_x (-1)^xJ(x)
    \label{eq:Sq}
\end{split}
\end{equation} 
Because the phase transition between the CDW and LE phases is a commensurate to incommensurate transition, when it is continuous, it should be in the Kosterlizt-Thouless university class. Therefore, in the CDW phase, we should see an antiferromagnetic pattern of pseudo-spin order and $J(k=\pi)\sim M^2L$ with M the order parameter approaching $\frac{1}{2}$ as $\omega_0$ decreases. And in the LE liquid phase, the spin-spin correlation should exhibits power-law behavior $J(k=\pi)\sim L^{1-\eta}$ where $\eta > 1/4$ such that $\eta \to 1/4$ upon approach to the transition point. In this spirit, we plot $J(k=\pi)/L$ for $L=100, 150, 300$. As shown in Fig.~\ref{fig:Sq_PT}, there is a clear crossing point at $\omega_0\approx 67$ for $J(k=\pi)/L^{3/4}$ with different L , which thus confirms the existence of a KT transition between the CDW and the LE phases in the strong-coupling limit.

\section{Other numerical results}
In the lower left corner of Fig.~\ref{fig:phase}, $\text{A}-\text{G}$ refer to a few calculations (not our own) by various numerical methods. The model at points A - ($U=0.6$, $\omega_0=0.5$),
C - $(1.2, 0.5)$,  D - $(1.62, 1.2)$, and E - $(1.62, 0.4)$, were studied using QMC (CT-INT method), A and C at a temperature such that $\beta t$~=~50,  while for D and E  $\beta t=20$ \cite{Assaad}.  On the basis of these studies, it was inferred that A is in the LE phase, while C, D, and E are in the CDW phase. On the basis of an early DMRG study, it was concluded that point F - $(2, 1)$ is in a CDW phase \cite{White}. The two remaining points, B - $(1.0, 0.5)$ and G - $(3, 5)$, were identified as quantum critical points using a stochastic series expansion (SSE) quantum Monte Carlo method\cite{Clay},  augmented by a finite-size scaling analysis. 

There are manifestly some discrepancies between the conclusions drawn on the basis of these different numerical studies. Similarly, the smooth dotted line for the phase boundary shown in Fig. \ref{fig:phase} is somewhat to the right of the optimal phase boundary one might draw on the basis of the earlier numerics. Due to the rather high temperature at which the QMC studies were conducted in comparison to the theoretically expected  exponentially small CDW gap, we think that while these results may be qualitatively right, it  should be expected that they will not be quantitatively precise.  In any case, it is presently unclear if the detailed shape of this phase boundary should be adjusted to better accommodate the results of contemporary numerical studies, or if one should stick to the present smooth interpolation and attribute the discrepancies to numerical uncertainty.

\section{Discussions on the phases of the doped system}
Slightly away from half filling, it is likely that there is a single LE phase everywhere in the phase diagram.  The spin-gap that characterizes both phases of the half-filled system is expected to extend smoothly to the lightly doped system. On the other hand, the generalized Luttinger's theorem insures that for an incommensurate electron density, there must be a gapless mode at $2k_F$.  Thus, the only plausible phase is a LE liquid with a spin-gap and power law CDW correlations.

There is one subtlety here worth noting.   For $\omega_0 = 0$, slight doping is expected to produce a state consisting of an array of solitons or discommensurations.\cite{HKSSReview}  These will produce mid-gap states, resulting in a spin-gap that is half the value of the spin-gap in the undoped system.  Upon including quantum fluctuations (i.e. for small but non-zero $\omega_0$) the soliton lattice will melt to form a power-law phase with $K_c =2$ (corresponding to dilute hard-core bosons or  spinless fermions), but the spin-gap is expected to be largely unaffected.

It is also possible that at larger deviations from half-filling, CDW order with higher order commensurability - for example for the 1/3 filled band - can arise, especially in the small $\omega_0$ limit.

\section{Conclusions}  Our major finding is the phase diagram in Fig. 1.  The topology of the phase diagram rests on general arguments, although the possibility of additional phases at intermediate $U/t$ and $\omega_0/t$ has not been definitively excluded.  Moreover, the asymptotic forms of the phase boundary in the upper and lower corners of the phase diagram have been supported by what we believe to be a convincing analysis.  The dotted part of the phase diagram is a sketch, drawn so as to smoothly connect with the established results in the asymptotic regimes.  The quantitative disagreements between this sketch and some of the earlier numerical results (indicated by the grey points in the figure) may either reflect some quantitative uncertainty in those results or may imply a more convoluted shape to the phase boundary.

The phase transition between the CDW and LE phases is a commensurate to incommensurate transition, so where it is continuous it should be described by a 1+1 dimensional sine-Gordon theory and should thus be in the Kosterliz-Thouless universality class. This has been verified by the strong coupling calculations in the upper right corner of the phase diagram. However, it is not precluded that it could be first order along other parts of its extent.

\subsection*{{Acknowledgement}}
We are grateful to Cheng Peng for helpful discussions on DMRG method. The DMRG calculations were performed using the ITensor Library \cite{itensor}. Part of the computational work was performed on the Sherlock cluster at Stanford. This work was supported in part by the U.S. Department of Energy (DOE), Office of Basic Energy Sciences, Division of Materials Sciences and Engineering (SZ), NSF grant No. DMR-2000987 at Stanford (SAK), and NSF Grant DMR-2038011 (IE).

\onecolumngrid
\subsection*{{Supplemental Material}}

\subsection*{A. Effective Hamiltonian in the anti-adiabatic limit}

In this Appendix, we provide details on the derivation of the effective Hamiltonian that provides the first corrections around the anti-adiabatic limit, $\omega_0 \to \infty$. We do this in two ways, first via a path integral technique and then with Hamiltonian methods.

\subsection*{Path integral approach}

In path integral language, the Euclidean action is
	\begin{align}
	S[\psi^\dag,\psi,v] &= \int_0^\beta d\tau \left\{  \sum_{ij} \psi_{i\sigma}^\dag [(\partial_\tau  + \mu)\delta_{ij}+ t_{ij}]\psi_{j\sigma}   + \sum_i \left[\frac M2(\partial_\tau v_i)^2 + \frac K2 v_i^2 \right]  + \alpha \sum_i v_i \psi^\dag_{i\sigma} \psi_{i\sigma} \right\} \\
	&= \sum_n \left\{  \sum_{ij} \psi_{i\sigma,n}^\dag [(i\omega_n + \mu)\delta_{ij}+ t_{ij}]\psi_{j\sigma,n}   + \frac 12 \sum_i v_{i,-n} (M\nu_n^2 + K ) v_{i,n} + \alpha \sum_i v_{i,-n} \rho_{i\sigma,n} \right\}. 
	\end{align}
In the second line we transform to Matsubara frequencies $\omega_n = (2n+1)\pi/\beta$, $\nu_n = 2n\pi/\beta$ and we define the density $\rho_{i\sigma} = \psi^\dag_{i\sigma} \psi_{i\sigma}$. The phonon Green's function is	
    \begin{equation}
	D(\nu_n)  = \frac{1}{M\nu_n^2 + K} = \frac{1}{K} \frac{\omega_0^2}{\nu_n^2 + \omega_0^2} \to 
	\begin{cases} \delta_{\nu_n,0}/K & \omega_0 \to 0, \\
	1/K  & \omega_0 \to \infty \end{cases}
	\end{equation}
Integrating out the phonon fields yields a retarded electron-electron interaction:
	\begin{equation}
	S_\mathrm{int}[\psi^\dag,\psi] = -\frac{\alpha^2}{2} \sum_n  \sum_{i\sigma} \rho_{i\sigma,-n} D(\nu_n)  \rho_{i\sigma,n} = -\frac{\alpha^2}{2K} \sum_n  \sum_{i\sigma} \rho_{i\sigma,-n} \left(\frac{\omega_0^2}{\nu_n^2 + \omega_0^2}\right)  \rho_{i\sigma,n} 
	\end{equation}
For $\omega_0 = \infty$, the interaction is instantaneous and we recover the attractive Hubbard model with $U=\alpha^2/K$. We can expand around this limit in powers of $1/\omega_0$. This is equivalent to a gradient expansion in imaginary-time derivatives. The result is
	\begin{equation}
	S_\mathrm{int} = -\frac U2 \sum_n  \sum_{i\sigma} \rho_{i\sigma,-n}  \rho_{i\sigma,n} + \frac{U}{2\omega_0^2} \sum_n  \sum_{i\sigma} \rho_{i\sigma,-n} \nu_n^2\rho_{i\sigma,n}   + \mathcal O\left(\frac{1}{\omega_0^4}\right)
	\end{equation}
In imaginary time, the second term is 
	\begin{equation}
	S_\text{int}^{(2)}=-\frac{U}{2\omega_0^2} \int_0^\beta d\tau \sum_{i\sigma} (\partial_\tau \rho_{i\sigma})^2.
	\label{eq:Sint2}
	\end{equation}
	
\subsection*{Hamiltonian approach}	

For a Hamiltonian approach, consider a unitary transformation of the Hamiltonian
	\begin{equation}
	H' = UHU^\dag, \quad U = \prod_i e^{i\alpha p_in_i/K}
	\end{equation}
where $\rho_i = \sum_\sigma \rho_{i\sigma} = \sum_\sigma c^\dag_{i\sigma}c_{i\sigma}$. The result is 
	\begin{equation}
	H' = -\sum_{ij} t_{ij} e^{i\alpha(p_i - p_j)/K} c^\dag_{i\sigma} c_{j\sigma} - \frac U2 \sum_i \rho_i^2+ \sum_i \frac{p_i^2}{2M} + \frac K2 v_i^2.
	\end{equation}
The transformation removes the bilinear electron-phonon coupling, at the cost of introducing an attractive interaction electron-electron interaction and adding electron-phonon interaction into the hopping matrix elements. To find an expansion around $\omega_0 = \infty$, rewrite the phonon coordinates and conjugate momenta in terms of the creation and annihilation operators:
	\begin{equation}
	p = i \sqrt{M\omega_0/2} (b^\dag - b) \quad \Rightarrow \quad e^{i\alpha p /K} = e^{-\sqrt{U/(2\omega_0)}(b^\dag - b)},
	\end{equation}
so that we may expand
	\begin{equation}
	H' \approx -\sum_{ij} t_{ij} \left\{ 1 + \sqrt{\frac{U}{2\omega_0}} [(b_j - b_j^\dag) - (b_i - b_i^\dag)]\right\} c^\dag_{i\sigma} c_{j\sigma} - \frac U2 \sum_i \rho_i^2+ \omega_0 \sum_i (b_i^\dag b_i + 1/2).
	\end{equation}
Specializing to the case of nearest-neighbor hopping, this expansion yields a coupling between the conjugate momentum of the phonon and  the ``lattice divergence" of the current: 
	\begin{equation}
	\tilde H_\mathrm{int}^{(2)} = \sqrt{\frac{U}{2\omega_0}}\sum_{n\sigma} i(b_n - b^\dag_n)  (j_n - j_{n-1}),
	\end{equation}
where the local current operator is 
	\begin{equation}
	j_n = it (c^\dag_n c_{n+1} - c^\dag_{n+1}c_n).
	\end{equation}
In momentum space, 
	\begin{equation}
	\tilde H_\mathrm{int}^{(2)} = \sqrt{\frac{U}{2N\omega_0}}t\sum_{kq\sigma} f_{kq} (b_q - b^\dag_{-q}) c^\dag_{k+q \sigma} c_{k\sigma}
	\end{equation}
where
	\begin{equation}
	f_{kq} = -2i [\cos(k+q) - \cos k].
	\end{equation}
Direct perturbation theory about the $\omega_0=\infty$ limits yields the effective electron-electron interaction:
	\begin{equation}
	H_\mathrm{int}^\mathrm{(2)} = -\frac UN \left(\frac{t}{\omega_0}\right)^2 \sum_{kq} V_{kk'q}  c^\dag_{k+q\sigma}c_{k\sigma} c^\dag_{k'-q \sigma'} c_{k'\sigma'},
	\label{eq:Hint_eff}
	\end{equation}
where 
	\begin{equation}
	V_{kk'q} = f_{k,q}f_{k',-q} = -4(1-\cos q) [\cos(k+k') - \cos(k-k'+q)]
	\end{equation}
The continuity equation relates Eqs.~\eqref{eq:Hint_eff} and \eqref{eq:Sint2}.

\subsection*{B. The adiabatic limit $\omega_0\to 0 $ (Derivation for the TLM model)}
\noindent Takayama, Lin-Liu, and Maki (TLM) have found a remarkable analytic solution for solitons in a condensed CDW system and their model is a continuum version of the SSH model \cite{TLM,HKSSReview,Horovitz}. With similar treatment, we find in the continuum limit, the effective field theory of the Holstein model is also the TLM model. Here is a short derivation.
The Holstein model is defined as:
\begin{equation}
    H=-t\sum_{n,\sigma}(c^\dag_{n\sigma}C_{n+1,\sigma}+H.c.)-\lambda\sum_{n,\sigma} x_nn_{n\sigma}+\frac{1}{2}K\sum_nx^2_n+\frac{1}{2}M\sum_n \dot{x_n}^2
\end{equation}

\noindent Let
\begin{equation}
    x_n=e^{i\pi n}z_n
    \label{eq:x_n}
\end{equation}
Then the coupling term becomes:
\begin{equation}
\begin{split}
    \lambda\sum_{n,\sigma}(-1)^nz_nc^\dag_{n,\sigma}c_{n,\sigma}&=\lambda \sum_{n,\sigma}\sum_{kqq'} z_k\ e^{ikn}e^{i\pi n}e^{iqn}e^{-iq'n}c^\dag_{q,\sigma}c_{q',\sigma} \\[2ex]
    &=\lambda \sum_{kqq',\sigma}z_k\ c^\dag_{q,\sigma}c_{q',\sigma} \delta_{k+\pi+q-q'}\\[2ex]
    &=\lambda\sum_{kq,\sigma}z_k\ c^\dag_{q,\sigma}c_{\pi+k+q,\sigma}
\end{split}
\end{equation}
Now let $p=q-\frac{\pi}{2}$ and $p'=q'+\frac{\pi}{2}$
\begin{equation}
\begin{split}
    \lambda\sum_{kq,\sigma}z_k\ c^\dag_{q,\sigma}c_{\pi+k+q,\sigma}&=\lambda\sum_{k,|p|<\frac{\pi}{2},\sigma}z_k\ c^\dag_{k_F+p,\sigma}c_{-k_F+k+p,\sigma}+\lambda\sum_{k,|p|<\frac{\pi}{2},\sigma}z_k\ c^\dag_{-k_F+p,\sigma}c_{k_F+k+p,\sigma}\\[2ex]
    &= \int dx\ \lambda z(x)\Big[L_{\sigma}^\dag(x) R_{\sigma}(x)+R_{\sigma}^
    \dag(x) L_{\sigma}(x)\Big]
\end{split}
\end{equation}

\noindent The free fermion part is (set $\hbar=v_F=1)$: 
\begin{equation}
    H_0=\sum_{\sigma}\int dx -(R_{\sigma}^\dag i \partial_xR_{\sigma}-L_{\sigma}^\dag i \partial_x L_{\sigma})=\sum_{\sigma}\int dx\  \psi_{\sigma}^\dag (x)[-i\sigma_z\partial_x] \psi_{\sigma}(x)
\end{equation}
where $\psi_{\sigma}(x)$ is a spinor made up of the right-moving $R_{\sigma}(x)$ and left-moving $L_{\sigma}(x)$ components of the Fermi field near the Fermi points. 
\\And the free phonon part is:
\begin{equation}
\begin{split}
    \frac{1}{2}K\sum_n x^2_n+\frac{1}{2}M\sum_n\dot{x}^2_n=\frac{1}{2}K\sum_n e^{i2\pi n}z^2_n+\frac{1}{2}M\sum_ne^{i2\pi n}\dot{z}^2_n
\end{split}
\end{equation}
So in the continuum limit, the Holstein model is also the TLM model:
\begin{equation}
\begin{split}
    H=\int dx\  \psi^\dag (x)[-i\sigma_z\partial_x] \psi(x)\ +\ \lambda z(x)\psi^\dag(x)\sigma_x\psi(x)+\int dx\ K\Big[\frac{\dot{z}(x)^2}{\omega^2_0}+z^2(x)\Big]
\end{split}
\end{equation}

\subsection*{C. The strong-coupling $U\to\infty$ limit}
The strong-coupling expansion for the 1D Holstein model can be schematically expressed as the following diagrams. (a) denotes the hopping of an electron from site i to site j then back to site i, which is the only possibility for the second-order term. Similarly, (b),(c) represent two possible fourth-order processes while the unlinked diagram is not included here since its contributions vanish.
\begin{figure}[h]
    \centering
  \includegraphics[width=0.6\linewidth]{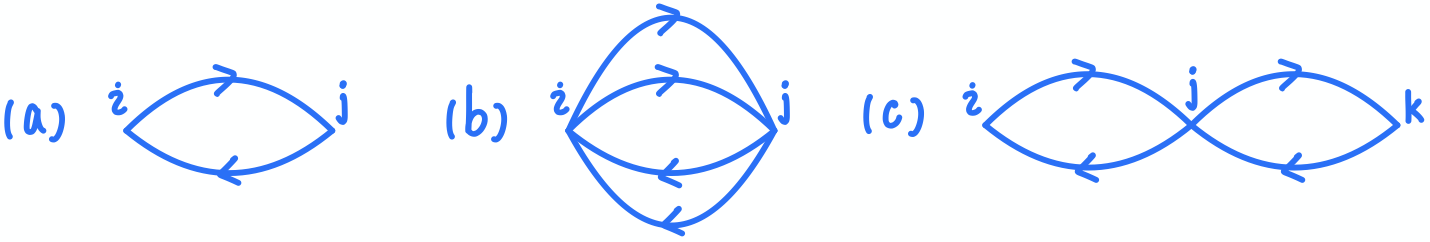}
    \caption{The second-order and fourth-order diagrams which is used in the determination of the effective Hamiltonian. This figure is reproduced from Ref.~\cite{strong}.}
\end{figure}
\\Then with pseudospin operators defined as Eq.\eqref{eq:pseudospin}, the corresponding terms in the effective Hamiltonian are \cite{strong}: 
\begin{equation}
\begin{split}
    H^{(2)}=\frac{1}{2}\sum_{i}\Bigg[j^{(2)}_{\perp}(i)\frac{1}{2}\Big(J^+_iJ^-_{i+1}+J^-_iJ^+_{i+1}\Big) +j^{(2)}_{\parallel}(i)\Big(J^z_iJ^z_{i+1}-\frac{1}{4}\Big)\Bigg]
\end{split}
\end{equation}
\begin{equation}
\begin{split}
    H^{(4)}=&\frac{1}{2}\sum_{i}\Bigg[\Big(j^{(4)}_{\perp}(i)+j'_{\perp}(i)\Big)\frac{1}{2}\Big(J^+_iJ^-_{i+1}+J^-_iJ^+_{i+1}\Big)+j''_{\perp}(i)\frac{1}{2}\Big(J^+_iJ^-_{i+2}+J^-_iJ^+_{i+2}\Big) \\[1ex]
    & +\Big(j^{(4)}_{\parallel}(i)-j'_{\parallel}(i)\Big)\Big(J^z_iJ^z_{i+1}-\frac{1}{4}\Big)+\Big(j'_{\parallel}(i)+j''_{\parallel}(i)\Big)\Big(J^z_iJ^z_{i+2}-\frac{1}{4}\Big)\Bigg]\\[2ex]
\end{split}
\end{equation}
where the explicit expressions for eight coefficients are \cite{strong}: 
\begin{equation}
\begin{split}
&j^{(2)}_{\perp}=-2\Big(-\frac{2t^2}{|U|}e^{-2S}\Big)\Big(1+\sum^{\infty}_{n=1}\frac{S^n}{(S+1)(S+2)...(S+n)}\Big)\\[2ex]
&j^{(2)}_{\parallel}=-2\Big(-\frac{2t^2}{|U|}\Big)\Big(1+\sum^{\infty}_{n=1}\frac{(-S)^n}{(S+1)(S+2)...(S+n)}\Big)
\end{split}
\end{equation}

\begin{equation}
\begin{split}
j^{(4)}_{\perp}=\frac{8t^4}{|U|^3}S^3e^{-2S} &\Bigg[\sum^{\infty}_{\substack{m,m'=0\\m+m'\neq 0}}\int^1_0 dx\int^1_0dy \ (xy)^{S-1}\ 2\text{cosh}[S(x-y)] \ \frac{(S/2)^{m+m'}(1-x^2)^m(1-y^2)^{m'}}{m!m'!(m+m')}\\[2ex]
&-\sum^{\infty}_{m,m'=0}
\frac{S^{m+m'}}{m!m'!}\frac{(-1)^m+(-1)^{m'}}{(m+S)^2(m'+S)}\Bigg]
\end{split}
\end{equation}

\begin{equation}
\begin{split}
j^{(4)}_{\parallel}=\frac{8t^4}{|U|^3}S^3e^{-2S}\Bigg[&\sum^{\infty}_{\substack{m,m'=0\\m+m'\neq 0}}\int^1_0 dx\int^1_0dy \ (xy)^{S-1}\Bigg(e^{S(x+y)}\frac{(S/2)^{m+m'}(1-x)^{2m}(1-y)^{2m'}}{m!m'!(m+m')}\\[2ex]
&+e^{-S(x+y)}\frac{(S/2)^{m+m'}(1+x)^{2m}(1+y)^{2m'}}{m!m'!(m+m')}\Bigg)-\sum^{\infty}_{m,m'=0}\frac{S^{m+m'}}{m!m'!}\frac{1+(-1)^{m+m'}}{(m+S)^2(m'+S)}\Bigg]
\end{split}
\end{equation}

\begin{equation}
\begin{split}
j'_{\perp}=\frac{4t^4}{|U|^3}S^3e^{-2S}\Bigg[&\int^1_0 dx\int^1_0dy \int^1_0 dz (xyz)^{S-1}\Bigg(\text{exp}\Big\{\frac{S}{2}[x-y+z-2z(x-y)-xyz]\Big\}+\text{exp}\Big\{\frac{S}{2}[x-y-z(x+y)]\Big\}\Bigg)\\[2ex]
&+2\int^1_0 dx\int^1_0 dy(xy)^{S-1}e^{S(x-y)}\sum^{\infty}_{m=1}\frac{(S/2)^m(1-x)^m(1+y)^m}{m!m}+\int^1_0 dx\int^1_0 dy (xy)^{S-1}e^{S(x-y)}(\text{lnx}+\text{lny})\Bigg]
\end{split}
\end{equation}

\begin{equation}
\begin{split}
j''_{\perp}=-\frac{4t^4}{|U|^3}S^3e^{-2S}\Bigg[&\int^1_0 dx\int^1_0dy \int^1_0 dz (xyz)^{S-1} \text{exp}\Big\{\frac{1}{2}S[x+y+z-2z(x+y)+xyz]\Big\}\\[2ex]
&+2\int^1_0 dx\int^1_0 dy(xy)^{S-1}e^{-S(x+y)}\sum^{\infty}_{m=1}\frac{(S/2)^m(1+x)^m(1+y)^m}{m!m}+2\int^1_0 dx\int^1_0 dy(xy)^{S-1}e^{-S(x+y)}\text{lnx}\Bigg]
\end{split}
\end{equation}

\begin{equation}
\begin{split}
j'_{\parallel}=-\frac{4t^4}{|U|^3}S^3e^{-2S}\Bigg[&\int^1_0 dx\int^1_0dy \int^1_0 dz (xyz)^{S-1}\Bigg(\text{exp}\Big\{\frac{S}{2}[-x-y+z+2z(x+y)+xyz]\Big\}\\[2ex]
&+\text{exp}\Big\{\frac{S}{2}[-x-y+2z+z(x+y)+2xyz]\Big\}\Bigg)\\[2ex]
&+2\int^1_0 dx\int^1_0 dy\ (xy)^{S-1}e^{S(x+y)}\sum^{\infty}_{m=1}\frac{(S/2)^m(1-x)^m(1-y)^m}{m!m}+2\int^1_0 dx\int^1_0 dy\ (xy)^{S-1}e^{S(x+y)}\text{lnx}\Bigg]
\end{split}
\end{equation}

\begin{equation}
\begin{split}
j''_{\parallel}=&\frac{4t^4}{|U|^3}S^3e^{-2S}\int^1_0dx\int^1_0dy\int^1_0dz\ (xyz)^{S-1} \text{exp}\Big\{\frac{1}{2}S[x+y+2z-z(x+y)+2xyz]\Big\}
\end{split}
\end{equation}
Here the combination $t$ and $U$ comes out as the overall energy scale. And the only tuning parameter is the dimensionless retardation factor $S\equiv U/\omega_0$. We evaluate the values of eight coefficients as functions of S as shown in Fig.~\ref{fig:all_j}. Then $t_1,~t_2,~V_1,~V_2$ can be determined through Eq.~\eqref{eq:tVj2} and are plotted in Fig.~\ref{fig:tV}. We see at finite $|U|$, as $S\to 0$, i.e. $\omega_0\to \infty$ (the Hubbard limit), the values of $t_1,~t_2,~V_1,~V_2$ match the analytic expressions given in Eq.~\eqref{eq:tVj}.
\begin{equation}
\begin{split}
    &t_1=\frac{1}{4}\Big(j^{(2)}_{\perp}(i)+j^{(4)}_{\perp}(i)+j'_{\perp}(i)\Big)\ \xrightarrow{\omega_0 \to \infty}\ \frac{1}{4}\Big(\ \frac{4t^2}{|U|}-\frac{16t^4}{|U|^3}\ \Big)\\[1ex]
    &t_2=\frac{1}{4}j''_{\perp}(i)\ \xrightarrow{\omega_0 \to \infty}\ \frac{1}{4}\frac{4t^4}{|U|^3} \\[3ex]
    & V_1=\frac{1}{4}\Big(j^{(2)}_{\parallel}(i)+j^{(4)}_{\parallel}(i)-j'_{\parallel}(i)\Big)\ \xrightarrow{\omega_0 \to \infty}\ t_1 \\[3ex]
    &V_2=\frac{1}{4}\Big(j'_{\parallel}(i)+j''_{\parallel}(i)\Big)\ \xrightarrow{\omega_0 \to \infty}\ t_2
    \label{eq:tVj2}
\end{split}
\end{equation}
\begin{figure}[hbt!]
    \centering
  \includegraphics[width=1.0\linewidth]{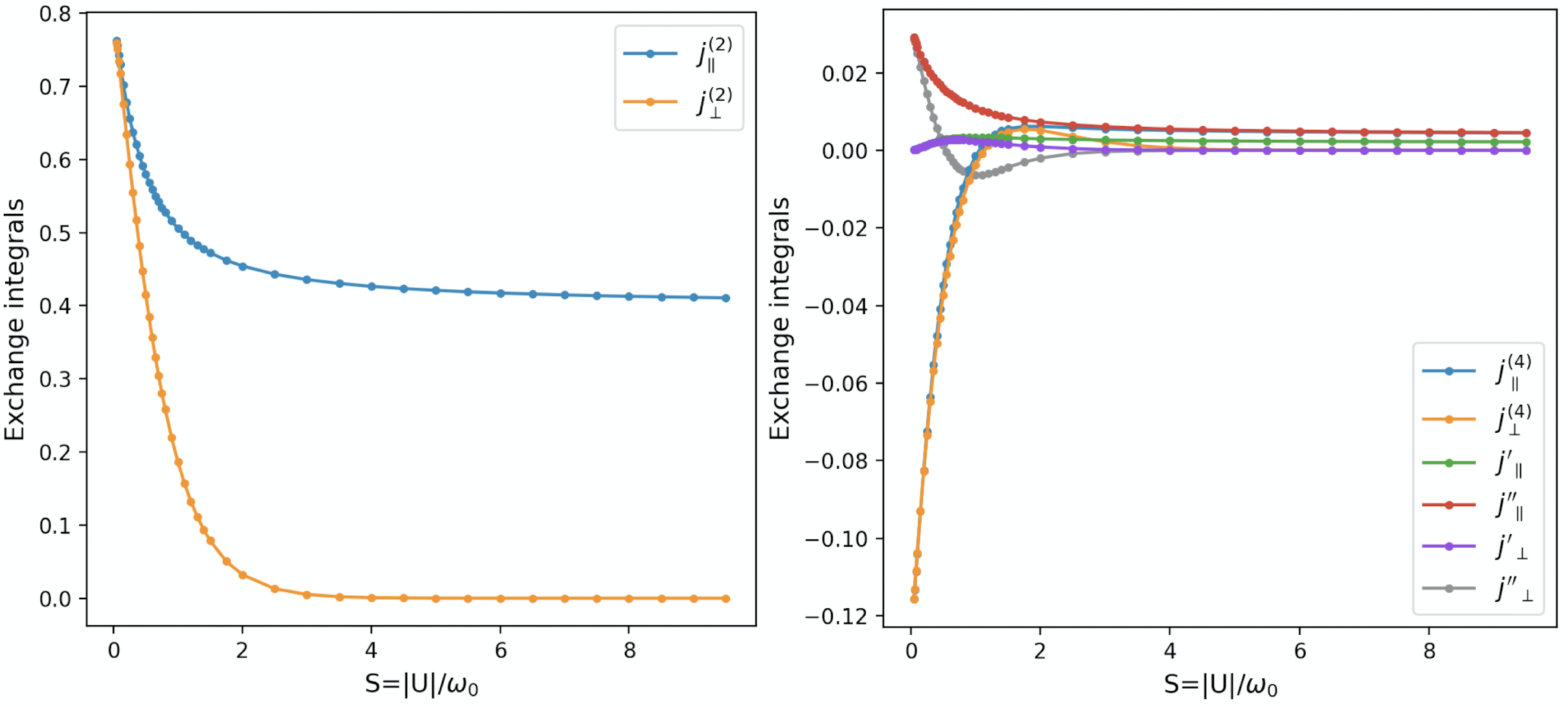}
  \caption{An illustration of the exchange integrals (coefficients of the effective pseudospin Hamiltonian) as functions of S with $t=1$. Here we use $|U|=5$ as an example.}
  \label{fig:all_j}
\end{figure}


\begin{thebibliography}{99}

\bibitem{Peierls} R. Peierls, Surprises in Theoretical Physics (Princeton University Press, New Jersey, 1979)

\bibitem{Pouget} J.-P. Pouget, C. R. Phys. 17, 332 (2016).

\bibitem{Hohenadler} M. Hohenadler and H. Fehske, The European Physical Journal B 91, 204 (2018).

\bibitem{Landau} L. D. Landau, Phys. Z. Sowjetunion 3, 644 (1933).

\bibitem{Holstein} T. Holstein, Ann. Phys. (N.Y.) 8, 325 (1959); 8, 343 (1959).

\bibitem{Alvermann} A. Alvermann, H. Fehske, and S. A. Trugman, Phys. Rev. B 81, 165113 (2010).

\bibitem{Assaad} J. Greitemann, S. Hesselmann, S. Wessel, F. F. Assaad,
and M. Hohenadler, Phys. Rev. B 92, 245132 (2015).

\bibitem{KM} K.-M. Tam, S.-W. Tsai, and D. K. Campbell, Phys. Rev. B 84,
165123 (2011).

\bibitem{Hohenadler2} M. Hohenadler and F. F. Assaad, Phys. Rev. B 87, 075149
(2013).

\bibitem{Clay} R. T. Clay and R. P. Hardikar, Phys. Rev. Lett. 95, 096401
(2005).

\bibitem{Hardikar} R. P. Hardikar and R. T. Clay, Phys. Rev. B 75, 245103 (2007).

\bibitem{Bakrim2} H. Bakrim and C. Bourbonnais, Phys. Rev. B 91, 085114
(2015).

\bibitem{Fehske} H. Fehske, G. Hager, and E. Jeckelmann, Europhys. Lett. 84, 57001 (2008).

\bibitem{Ejima} S. Ejima and H. Fehske, J. Phys.: Conf. Ser. 200, 012031 (2010).

\bibitem{White} E. Jeckelmann, C. Zhang, and S. R. White, Phys. Rev. B
60, 11, 7950 (1999).

\bibitem{Tezuka} M. Tezuka, R. Arita, and H. Aoki, Phys. Rev. B 76, 155114
(2007).

\bibitem{ED1} H. Fehske, G. Wellein, A. Weisse, F. Gohmann, H. Buttner, and A. R. Bishop, Physica B (Amsterdam) 312–313, 562 (2002).

\bibitem{ED2} H. Fehske, A. P. Kampf, M. Sekania, and G. Wellein, Eur. Phys. J. B 31, 11 (2003).

\bibitem{ED3} H. Fehske, G. Wellein, G. Hager, A. Weiße, and A. R. Bishop, Phys. Rev. B 69, 165115 (2004).

\bibitem{Bindloss} I. P. Bindloss, Phys. Rev. B 71, 205113 (2005).

\bibitem{Bakrim} H. Bakrim and C. Bourbonnais, Phys. Rev. B 76, 195115 (2007).

\bibitem{HF} J. E. Hirsch and E. Fradkin, Phys. Rev. B 27, 4302 (1983)

\bibitem{Bursill} R. J. Bursill, R. H. McKenzie, and C. J. Hamer, Phys. Rev. Lett. 80, 5607 (1998).

\bibitem{Wellein} M. Hohenadler, G. Wellein, A. R. Bishop, A. Alvermann, and H. Fehske, Phys. Rev. B 73, 245120 (2006).

\bibitem{Ejima2} S. Ejima and H. Fehske, Europhys. Lett. 87, 27001 (2009).

\bibitem{SSH} E. Fradkin and J. E. Hirsch, Phys. Rev. B 27, 1680 (1983).

\bibitem{Supplemental} See Supplemental Material below.

\bibitem{Giamarchi} T. Giamarchi, Quantum Physics in One Dimension (Clarendon, Oxford, 2004).

\bibitem{Voit} J. Voit, Rep. Prog. Phys. 58, 977 (1995).

\bibitem{entropy1} P. Calabrese and J. Cardy, Journal of Statistical Mechan- ics: Theory and Experiment 2004, P06002 (2004).

\bibitem{entropy2} M. Fagotti and P. Calabrese, Journal of Statistical Me- chanics: Theory and Experiment 2011, P01017 (2011).

\bibitem{TLM} H. Takayama, Y. R. Lin-Liu, and K. Maki, Phys. Rev. B 21, 2388 (1980).

\bibitem{HKSSReview} A. J. Heeger, S. Kivelson, J. R. Schrieffer, and W. P. Su, Rev. Mod. Phys., 60 (1988), p.781

\bibitem{NM1} M. Nakahara, and K. Maki, Phys. Rev. B 25, 7789 (1982).

\bibitem{NM2} M. Nakahara, and K. Maki, Synth. Met. 13, 149 (1986).

\bibitem{Horovitz} B. Horovitz, Phys. Rev. B 22, 1101 (1980).

\bibitem{strong} J. K. Freericks , Phys. Rev. B 48, 3881 (1993).

\bibitem{twostep} G. T. Zimanyi, S. A. Kivelson, and A. Luther, Phys. Rev. Lett. 60, 2089 (1988)

\bibitem{RG} L. G. Caron and C. Bourbonnais, Phys. Rev. B 29, 4230 (1984)

\bibitem{itensor} M. Fishman, S. R. White, and E. M. Stoudenmire, “The ITensor software library for tensor network calculations,” (2020), arXiv:2007.14822.


\end{thebibliography}
\end{document}